\documentclass[pdflatex,sn-nature]{sn-jnl}% Style for submissions to Nature Portfolio journals
\usepackage{graphicx}%
\usepackage{multirow}%
\usepackage{amsmath,amssymb,amsfonts}%
\usepackage{amsthm}%
\usepackage{mathrsfs}%
\usepackage[title]{appendix}%
\usepackage{multirow}
\usepackage{textcomp}%
\usepackage{manyfoot}%
\usepackage{booktabs}%
\usepackage{algorithm}%
\usepackage{algorithmicx}%
\usepackage{algpseudocode}%
\usepackage{listings}%
\usepackage{gensymb}
\usepackage{textcomp}
\usepackage[table]{xcolor}
\usepackage{colortbl}
\usepackage[normalem]{ulem}
\usepackage{xcolor,colortbl,booktabs}

\definecolor{sigOneTwo}{RGB}{114,188,213}   % 1--2 sigma
\definecolor{sigTwoThree}{RGB}{255, 208, 111}  % 2--3 sigma
\definecolor{sigOut}{RGB}{231,98,84}        % >3 sigma
\theoremstyle{thmstyleone}%
\theoremstyle{thmstyletwo}%

\theoremstyle{thmstylethree}%

\newcommand{\EDFig}[1]{Extended Data Fig.~\ref{#1}}
\newcommand{\EDTable}[1]{Extended Data Table.~\ref{#1}}
\newcommand{\EDSec}[1]{Method Sec.~\ref{#1}}
\newcommand{\EDEq}[1]{Eq.~(\ref{#1})}
\renewcommand{\thesubsection}{\alph{subsection}}
\newcommand{\RefCite}[1]{ref.~\cite{#1}}

\begin{document}
\title[Article Title]{Flux-ratio anomalies in cusp quasars reveal dark matter beyond CDM}
%\title[Article Title]{Probing dark matter microphysics with cusp flux-ratio anomalies}

%%=============================================================%%
%% GivenName	-> \fnm{Joergen W.}
%% Particle	-> \spfx{van der} -> surname prefix
%% FamilyName	-> \sur{Ploeg}
%% Suffix	-> \sfx{IV}
%% \author*[1,2]{\fnm{Joergen W.} \spfx{van der} \sur{Ploeg} 
%%  \sfx{IV}}\email{iauthor@gmail.com}
%%=============================================================%%

\author[1,2]{\fnm{Siyuan} \sur{Hou}}%\email{syhou@pmo.ac.cn}

\author[1,2]{\fnm{Shucheng} \sur{Xiang}}

\author*[1,2]{\fnm{Yue-Lin Sming} \sur{Tsai}}\email{smingtsai@pmo.ac.cn}

\author*[1,2]{\fnm{Daneng} \sur{Yang}}\email{yangdn@pmo.ac.cn}

\author[1,2]{\fnm{Yiping} \sur{Shu}}

\author[3]{\fnm{Nan} \sur{Li}}

\author[1,4,5]{\fnm{Jiang} \sur{Dong}}

\author[1,6,7]{\fnm{Zizhao} \sur{He}}

\author*[1,2]{\fnm{Guoliang} \sur{Li}}\email{guoliang@pmo.ac.cn}

\author*[1,2]{\fnm{Yizhong} \sur{Fan}}\email{yzfan@pmo.ac.cn}

\affil[1]{\small\orgdiv{Purple Mountain Observatory}, \orgname{Chinese Academy of Sciences}, \orgaddress{\city{Nanjing}, \postcode{210023}, \state{Jiangsu}, \country{China}}}

\affil[2]{\small\orgdiv{School of Astronomy and Space Sciences}, \orgname{University of Science and Technology of China}, \orgaddress{\city{Hefei}, \postcode{230026}, \state{Anhui}, \country{China}}}

\affil[3]{\small\orgdiv{National Astronomical Observatories}, \orgname{Chinese Academy of Sciences}, \orgaddress{\city{Beijing}, \postcode{100101}, \state{Beijing}, \country{China}}}

\affil[4]{\small\orgdiv{Institute for Frontier in Astronomy and Astrophysics}, 
\orgname{Beijing Normal University}, 
\orgaddress{\city{Beijing}, \postcode{102206}, \country{China}}}

\affil[5]{\small\orgdiv{School of Physics and Astronomy}, 
\orgname{Beijing Normal University}, 
\orgaddress{\city{Beijing}, \postcode{100875}, \country{China}}}

\affil[6]{\small\orgdiv{Department of Physics}, \orgname{Nanchang University}, \orgaddress{\city{Nanchang}, \postcode{330031}, \state{Jiangxi}, \country{China}}}

\affil[7]{\small\orgdiv{Center for Relativistic Astrophysics and High Energy Physics}, \orgname{Nanchang University}, \orgaddress{\city{Nanchang}, \postcode{330031}, \state{Jiangxi}, \country{China}}}

%%==================================%%
%% Sample for unstructured abstract %%
%%==================================%%

%\abstract{Strongly lensed quasars in cusp configurations provide a uniquely sensitive probe of small-scale dark matter structure. 
%Yet multipole structure originating from disk components, asymmetric stellar profiles or merger remnants can mimic these anomalies and obscure their physical origin.
%Using the largest microlensing-free flux ratios for 17 quadruply imaged cusps, combined with extensive Monte Carlo simulations of mock lens realizations under cold dark matter (CDM), self-interacting dark matter (SIDM), and fuzzy dark matter scenarios (FDM), we identify a region in which the observed flux-ratio anomalies persist even after allowing globally parameterized models (“macromodels”) with multipole freedom.
%Building on this, we propose a diagnostic framework that identifies narrow opening angle cusps as macromodel-independent probes of the nature of dark matter. Even with full multipole freedom, J1042+1641 is $>3\sigma$ incompatible with both CDM and SIDM.
%Our results yield a Bayes factor exceeding $100$, providing very strong evidence for FDM over even the most optimistic CDM and SIDM scenarios. Future microlensing-free flux ratio data can decisively confirm these findings. 
% Future microlensing-free flux ratio measurements will enable a broader assessment of the dark matter interpretation. }

\abstract{Strongly lensed quasars in cusp configurations provide a uniquely sensitive probe of small-scale dark matter structure. 
Using the largest microlensing-free flux ratios for 17 quadruply imaged cusps, we combine these with extensive Monte Carlo simulations of mock lens realizations under cold dark matter (CDM), self-interacting dark matter (SIDM), and fuzzy dark matter (FDM) scenarios.
Building on this, we propose a region (minor-axis and narrow major-axis cusp lenses) where flux-ratio anomalies persist even under globally parameterized models ("macromodels") with multipole freedom (capturing disk, asymmetric, or merger-driven structures).
Within this region, J1042+1641 is $>3\sigma$ incompatible with both CDM and SIDM.
Our results yield a Bayes factor exceeding $100$, providing very strong evidence for FDM over even the most optimistic CDM and SIDM scenarios. As only 11 cusp lenses lie within this region, extending to larger samples will be essential for assessing its statistical generality and for decisively confirming these findings with future microlensing-free flux ratio data.
}

%%================================%%
%% Sample for structured abstract %%
%%================================%%


%%\pacs[JEL Classification]{D8, H51}

%%\pacs[MSC Classification]{35A01, 65L10, 65L12, 65L20, 65L70}

\maketitle
% \tableofcontents

%Unveiling the nature of dark matter remains a central challenge in modern physics and cosmology. 
The fundamental nature of dark matter is one of the most important and long-standing questions in physics and cosmology. 
The cold dark matter (CDM) framework has achieved remarkable success on large scales, reproducing the cosmic web and the anisotropies of the cosmic microwave background with high fidelity~\cite{Springel:2005nw, Planck:2018vyg}. On galactic and sub-galactic scales, however, several challenges persist. Inner halo profiles and the diversity of dwarf-galaxy rotation curves both deviate from simple CDM expectations~\cite{Moore:1994yx, Oman:2015xda}, and discrepancies in small-scale clustering remain under active debate~\cite{Zhang:2025bju}.
Independent evidence from strong lensing also point to low-mass subhalos, where analyses of perturbed Einstein arcs tend to favor perturbers with unusually high concentrations~\cite{Minor:2020hic, Enzi:2024ygw, Li:2025kpb}. Moreover, observations reveal an apparent excess of galaxy--galaxy strong-lensing events in clusters~\cite{Meneghetti:2020yif, Tokayer:2024wwo}.
These challenges motivate dark matter scenarios beyond CDM, among which self-interacting dark matter (SIDM) and fuzzy dark matter (FDM) have emerged as compelling alternatives. 
Strong-lensing magnification and flux-ratio measurements can probe subhalos down to $\sim10^6~M_\odot$~\cite{Narayan:1996ba, Nierenberg:2014cga, Powell:2025rmj, Vegetti:2026mmx}, precisely the regime where CDM, SIDM, and FDM make distinct small-scale predictions. They therefore offer a uniquely powerful route to testing the nature of dark matter.

The cusp relation provides a stringent test of the smoothness of the lensing mass distribution~\cite{schneiderGravitationalLensEquation1992, Gaudi:2002hu}. When a source lies close to a cusp of the caustic in an otherwise smooth lens potential, three highly magnified images form near the critical curve.
When the source approaches a cusp of the caustic, 
the signed magnifications of the three merging images nearly cancel, yielding the theoretical expectation $R_{\rm cusp} \approx |\mu_A+\mu_B+\mu_C|/(|\mu_A|+|\mu_B|+|\mu_C|) \approx 0$. Here $\mu_A$, $\mu_B$, and $\mu_C$ denote the signed magnifications of the three images that merge as the source approaches the cusp.
Since the images form extremely close to the critical curve, $R_{\rm cusp}$ becomes highly sensitive to perturbations in the lens potential~\cite{Mao:1997ek, Metcalf:2001ap}, and this sensitivity grows as the opening angle $\phi$ decreases, thereby strengthening its response to small-scale dark matter structure.
In practice, $R_{\rm cusp}$ is an observational, macromodel-independent statistical quantity constructed directly from image positions and flux ratios, while the opening angle $\phi$ provides a continuous geometric descriptor of the configuration.

Various dark matter scenarios invoke different mechanisms to violate the cusp relation. 
In the CDM picture, subhalos and halos along the line of sight perturb the lensing potential and can drive departures from the ideal cusp relation~\cite{Xu:2009ch}. 
However, the associated convergence fluctuations remain modest, 
making it difficult for CDM substructure alone to account for the most extreme anomalies~\cite{Xu:2014dda, Gilman:2016uit}.
SIDM offers a promising framework for addressing these anomalies by producing a population of dense, core-collapsed subhalos~\cite{Girmohanta:2022dog}. 
In SIDM, elastic scattering between dark matter particles drives gravothermal evolution~\cite{Spergel:1999mh, Tulin:2017ara, Adhikari:2022sbh, Hou:2025gmv}. 
Halos first develop central cores and subsequently undergo gravothermal collapse, 
producing compact high-density subhalos that generate pronounced flux-ratio anomalies and violations of the cusp relation. 
Velocity-dependent cross sections can nevertheless drive collapse in systems as low as $10^6~M_\odot$~\cite{Yang:2022hkm}, even for cross sections small enough to remain consistent with dwarf galaxies~\cite{Zhang:2025bju}.
By contrast, FDM suppresses bound substructure and instead introduces wave-induced density fluctuations that perturb the lens mapping~\cite{Hui:2021tkt, Kunkel:2022ldl, Schive:2025bcm}. 
Alfred et al.~\cite{Amruth:2023xqj} used high-precision radio observations to interpret the anomalies in HS~0810+2554, 
demonstrating that interference-induced fluctuations from FDM can significantly perturb the image plane even in the absence of traditional subhalos.

Deviations from the cusp relation have long been used to probe small-scale dark matter structure, 
but their constraining power among different opening angle configurations has not been systematically quantified. 
Higher-order multipoles in the lens galaxy, arising from stellar discs, mergers, and boxy or disky morphologies~\cite{Evans:2002ck, Keeton:2002qt, Kawano:2004sr, Congdon:2005iw}, can generate magnification perturbations that mimic signals otherwise attributed to dark matter substructure~\cite{Shan:2025glh}.
However, their dependence on lens geometry remains poorly characterized, and the extent to which they are degenerate with the microphysical predictions of CDM, SIDM, and FDM has not been assessed in a statistical framework. 
Previous analyses~\cite{Liu:2025hjj} have generally relied on flux ratios affected by stellar microlensing and have been limited by small sample sizes with narrow opening angle coverage, reducing their ability to distinguish between dark matter scenarios.

In this work, we use the latest microlensing-free flux ratios for 17 quadruply imaged cusp lenses (see \EDTable{tab:cusp_sample_obs} for details), drawn from the \textit{GO-2046 JWST} program~\cite{Nierenberg:2023tvi,Keeley:2024brx,Keeley:2025oig,Gilman:2025fhy} along with radio and HST narrow-line measurements.
Combining these data with extensive Monte Carlo simulations of mock lens realizations, 
we first construct a baseline set of lensing mock data based on singular isothermal ellipsoid (SIE) with external shear models~\cite{Yue:2022lcc, Dong:2024sij}, as well as SIE with external shear and higher-order multipoles models~\cite{Oh:2024dlj, Paugnat:2025bdn}.
The distributions of the macromodel parameters are chosen to be consistent with the observed lens sample. We explicitly compare the mock and observed distributions of the lens and source redshifts, the Einstein radius, the external shear, and the axis ratio (see \EDFig{fig:mock_sie_dist} for details).

We then superimpose substructure perturbations based on CDM, SIDM, and FDM scenarios using \texttt{pyHalo}\footnote{\url{https://github.com/dangilman/pyHalo}}~\cite{Gilman:2019nap} on the smooth lens potentials.
Using our \texttt{JAX}-accelerated GPU implementation\footnote{\url{https://github.com/jax-ml/jax}} 
of the \texttt{PICS} lensing framework~\cite{Li:2015tmf,jax2018github}, 
we generate $10^{8}$ cusp configurations for Bayesian model comparison.
This large ensemble allows us to identify the regions in the $R_{\rm cusp}$--$\phi$ plane where higher-order multipole perturbations dominate, and 
to assess how the behavior of several dark matter benchmark scenarios differs (see Methods for details).

%%%%%%%%%%%%%%%%%%%%%%%%%%%%%%%%%%%%%%%%%%%%%%%%%%%%%%%%%%%%%%%%%%%%%%%
\section*{Substructure perturbations in CDM, SIDM, and FDM scenarios}
%%%%%%%%%%%%%%%%%%%%%%%%%%%%%%%%%%%%%%%%%%%%%%%%%%%%%%%%%%%%%%%%%%%%%%%

\begin{figure}
    \centering
    \includegraphics[width=\linewidth]{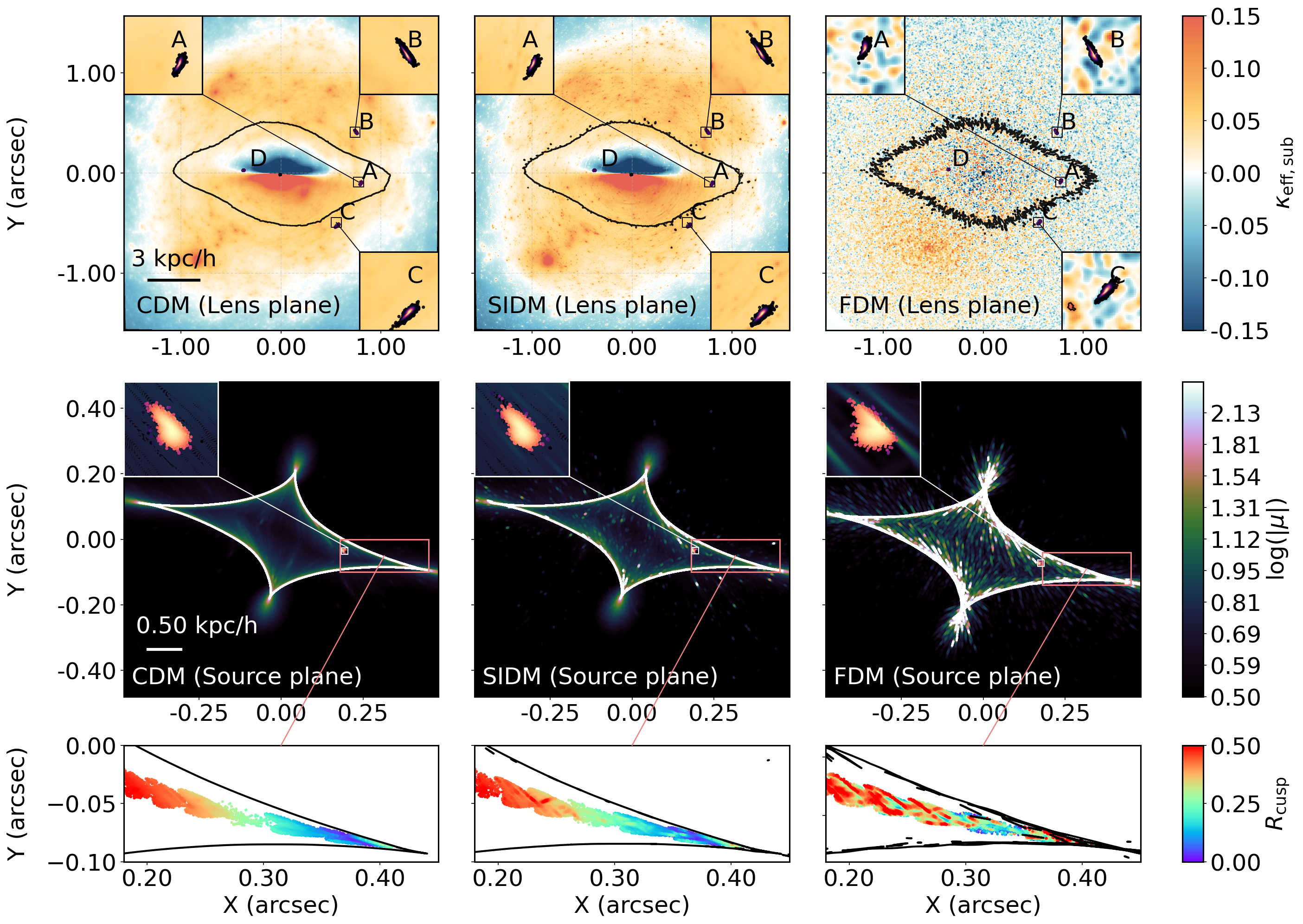}
    \caption{
    \textbf{Top panels:} Effective convergence perturbations ($\kappa_{\mathrm{eff,sub}}$; see \EDEq{eq:kappa_eff_sub}) induced by substructure in a representative lens with significant $m=3/4$ multipoles, characterized by a major-axis angle $\phi = 70.63^\circ$ (see \EDTable{tab:mock_all_sim_185_key} for macromodel parameters). The black ellipse denotes the critical curve, and the insets show the distinct perturbation patterns associated with different dark matter scenarios at each cusp image position (A, B and C).
    The colours encode the angular misfit $\chi^{2}$ (see \EDEq{Eq:kasquare}), constructed from $\phi$ and $\phi_{1}$.
    \textbf{Middle panels:} Source-plane caustics, with background colour indicating $\log|\mu|$. Insets illustrate the corresponding source positions for the points in the top panels, with colours also encoding $\chi^{2}$. They highlight substructure-induced distortions and local variations in the caustic shape across different dark matter scenarios.
    \textbf{Bottom panels:} Scatter distributions of $R_{\mathrm{cusp}}$ obtained while traversing $\phi$ near the cusp angle. As the source approaches the narrow cusp, the differences among CDM, SIDM, and FDM increase substantially. In particular, interference-driven density fluctuations in FDM produce a larger fraction of high-$R_{\mathrm{cusp}}$ points.
    }
    \label{fig:substructure_maps}
\end{figure}

Lensing signatures produced by small-scale perturbations in the lensing potential provide a sensitive probe of the nature of dark matter. In both the CDM and SIDM scenarios, these perturbations arise from dark matter substructure, including subhalos within the macrolens and halos along the line of sight (LOS). In contrast, in the FDM scenario the dominant lensing perturbations are generated by wave-induced density fluctuations, such as a central solitonic core and granule-like interference patterns, with subhalos and LOS halos also included but contributing at a subdominant level.
In Fig.~\ref{fig:substructure_maps}, we compare CDM, SIDM, and FDM realizations for the same simulated lens system. For the SIDM case, we adopt a model in which the scattering cross section follows a Rutherford-like form with strong velocity dependence~\cite{Feng:2009hw,Ibe:2009mk}. In this model, massive dwarf galaxies develop extended cores, whereas halos with masses below $10^8~M_\odot$ predominantly undergo core collapse. For the FDM scenario, we assume a particle mass of $\sim10^{-22}~\mathrm{eV}$ (see \EDSec{Sec:Dark Matter Substructure} for details).

The upper panels of Fig.~\ref{fig:substructure_maps} show the effective convergence perturbations, $\kappa_{\mathrm{eff,sub}}$, induced by these substructures.
$\kappa_{\mathrm{eff,sub}}$ is defined as the residual convergence after subtracting the macromodel contribution and removing the mean convergence of all subhalos and LOS halos (see \EDSec{Sec:Multi-plane lensing} for details).

In both CDM and SIDM, perturbations are dominated by subhalos in the main halo and LOS halos. 
CDM produces localized distortions in the lensing potential (left column). 
In contrast, SIDM (middle column) features a similar subhalo count but with rapid gravothermal collapse of low-mass halos, 
resulting in larger local perturbations. 
In the FDM case (right column), perturbations are dominated by granules, which create continuous alternating positive and negative structures. These granules typically correspond to masses of $10^{6}$ to $10^{7}\,M_\odot$, with their abundance surpassing that of CDM and SIDM subhalos in the same mass range (see zoom-ins in the upper panels of Fig.~\ref{fig:substructure_maps}).

These fluctuations are most pronounced near the host halo center, generating density variations around the critical curve (upper panels) that propagate into distortions in the source-plane caustics (middle panels). 
This enhances the sensitivity of $R_{\rm cusp}$ to FDM-induced perturbations, 
with FDM producing the largest effects, followed by SIDM, and CDM showing the weakest response. 
This also demonstrates that wave-like dark matter, even without bound subhalos, can produce significant lensing signatures.

To construct a well-sampled theoretical prediction map, we generate large ensembles of mock lens systems with dark matter substructures. 
Each realization is categorized by lens configurations with a specified opening angle $\phi$ through an MCMC search. 
Configurations are grouped into intervals of $\phi$ from $30^\circ$ to $150^\circ$ with a bin size of $10^\circ$. 
%We construct a kernel density estimate for each interval, independently normalized to remove the geometric imprint of the smooth lens model. This normalization ensures that changes in $R_{\mathrm{cusp}}$ reflect the impact of small-scale perturbations. 
For each $\phi$ bin, we construct a kernel density estimate (KDE) and normalize it independently. The resulting map captures the conditional scatter of $R_{\rm cusp}$ at fixed $\phi$, and therefore deviations from the smooth $R_{\rm cusp}$ distribution mainly reflect the impact of small-scale perturbations. (see \EDSec{sec:mcmc_sampling} for details).

%\section{Rcusp–$\phi$ sensitivity as a discriminator of dark matter microphysics}
%%%%%%%%%%%%%%%%%%%%%%%%%%%%%%%%%%%%%%%%%%%%%%%%%%%%%%%%%%%%%%%%%%%%%%%%%%%%%%%%%%%%%%%%
%\section*{Pure SIE with External Shear Case}
\section*{Statistical test with a SIE lens and shear}
%%%%%%%%%%%%%%%%%%%%%%%%%%%%%%%%%%%%%%%%%%%%%%%%%%%%%%%%%%%%%%%%%%%%%%%%%%%%%%%%%%%%%%%%

\begin{figure}
    \centering
    \includegraphics[width=\linewidth]{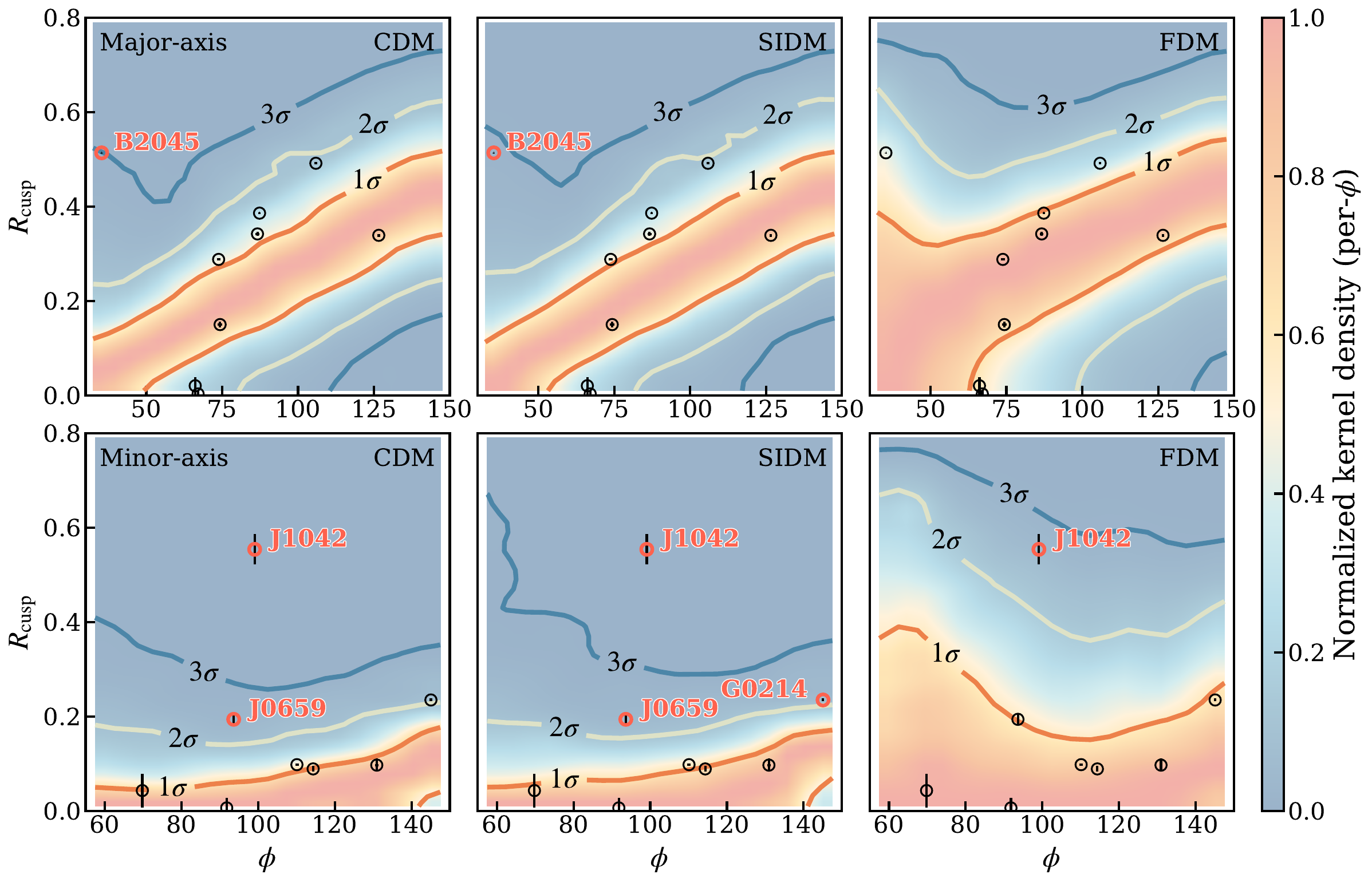}
    \caption{Normalized $R_{\mathrm{cusp}}$ distributions predicted under different dark matter scenarios for cusp-configured mock lens systems generated using a SIE model with external shear (see \EDFig{tab:Bey_smooth} for smooth-lens predictions). The lens populations are separated into major-axis and minor-axis cusp configurations (see \EDFig{fig:Major_Minor}). The distributions are constructed by samples $\phi$ from $30^\circ$ to $150^\circ$ and applying a normalized KDE within each $\phi$ bin. The $n\sigma$ contours correspond to highest-probability-density regions enclosing 68.3\%, 95\%, and 99.7\% of the total KDE probability mass, respectively. The left, middle, and right columns correspond to mock lenses populated with CDM, SIDM, and FDM substructures, respectively. Each dark matter system contains $\sim10^7$ data points. The background colour map represents the KDE-normalized density computed from these data points in each $\phi$ bin. Systems marked with red labels denote $2\sigma$ outliers (see \EDTable{tab:individual_norms} for the per-system density values).}
     \label{fig:NomonizedKDE}
\end{figure}

We compute the $R_{\rm cusp}$ distributions for mock lensing simulations under different dark matter scenarios and compare our results with those from 17 observed cusp-configured quadruply imaged quasars, as shown in Fig~\ref{fig:NomonizedKDE}. 
The simulated lenses naturally divide into two geometric categories: (i) major-axis cusp configurations follow a steeper $R_{\rm cusp}$--$\phi$ relation, while (ii) minor-axis cusp configurations lie on a shallower branch.
This difference arises because minor-axis cusp configurations place the source closer to the caustic cusp, 
which brings the three cusp images nearer to the critical curve for a fixed opening angle.
Consequently, minor-axis configurations are more sensitive to local, small-scale perturbations, while being insensitive to global distortions of the critical curve (such as large-scale ellipticity or higher-order multipoles). 
Hence, this geometric behavior allows us to discriminate between FDM and the CDM and SIDM scenarios.

The $R_{\rm cusp}$ distributions in Fig~\ref{fig:NomonizedKDE} show significant differences among dark matter scenarios, as visualized by KDE normalized densities in bins of $\phi$. 
In the SIDM case with gravothermal core collapse, which enhances small-scale structure, 
there is a modest shift in the predicted $R_{\rm cusp}$ distribution compared to CDM. 
However, this shift is insufficient to explain all observed outliers, leaving notable tensions. 
In contrast, FDM produces a distinct signature: for major-axis cusp configurations, only FDM generates a substantial population of narrow-cusp realizations that reach the high $R_{\rm cusp}$ values observed in B2045.

In Fig.~\ref{fig:NomonizedKDE}, the main differences appear between narrow major-axis cusp configurations and minor-axis cusp configurations.
In CDM and SIDM scenarios, the minor-axis cusps density is confined to a low-$R_{\rm cusp}$ ridge, 
while in FDM, it shifts systematically to higher $R_{\rm cusp}$ values, extending to $R_{\rm cusp} \sim 0.2-0.4$ within the $1\sigma$ band at fixed $\phi$. 
Systems lying outside the $2\sigma$ density band are extremely rare.
For major-axis cusp configurations, the observed system B2045 lies outside the $2\sigma$ region
in the CDM and SIDM KDE maps, while falling within the $2\sigma$ band predicted by FDM.
For minor-axis cusp configurations, J0659 and G0214 lie outside the $2\sigma$ density region in the CDM and SIDM KDE maps, but fall within the $1\sigma$ density band predicted by FDM, with J1042 remaining the lowest-$R_{\rm cusp}$ system.

We can quantify these trends using the Bayesian decision criterion (see details in \EDSec{Sec:Bey}). The Bayes-factor summary will be provided in Table~\ref{tab:Bey}. Smooth models without dark matter substructure are decisively rejected, with $\mathrm{BF} < 10^{-35}$ for all smooth prescriptions (See \EDTable{tab:Bey_smooth}). Among the SIE with external shear family of models, the evidence strongly favors FDM over SIDM and CDM, with $\mathrm{BF}(\mathrm{FDM}/\mathrm{CDM}) \sim 10^{6}$ and $\mathrm{BF}(\mathrm{FDM}/\mathrm{SIDM}) \sim 10^{4}$.

%%%%%%%%%%%%%%%%%%%%%%%%%%%%%%%%%%%%%%%%%%%
%\section*{Multipole Case}
\section*{Statistical test with a SIE lens, shear, and multipoles}
%%%%%%%%%%%%%%%%%%%%%%%%%%%%%%%%%%%%%%%%%%%

\begin{figure}
    \centering
    \includegraphics[width=\linewidth]{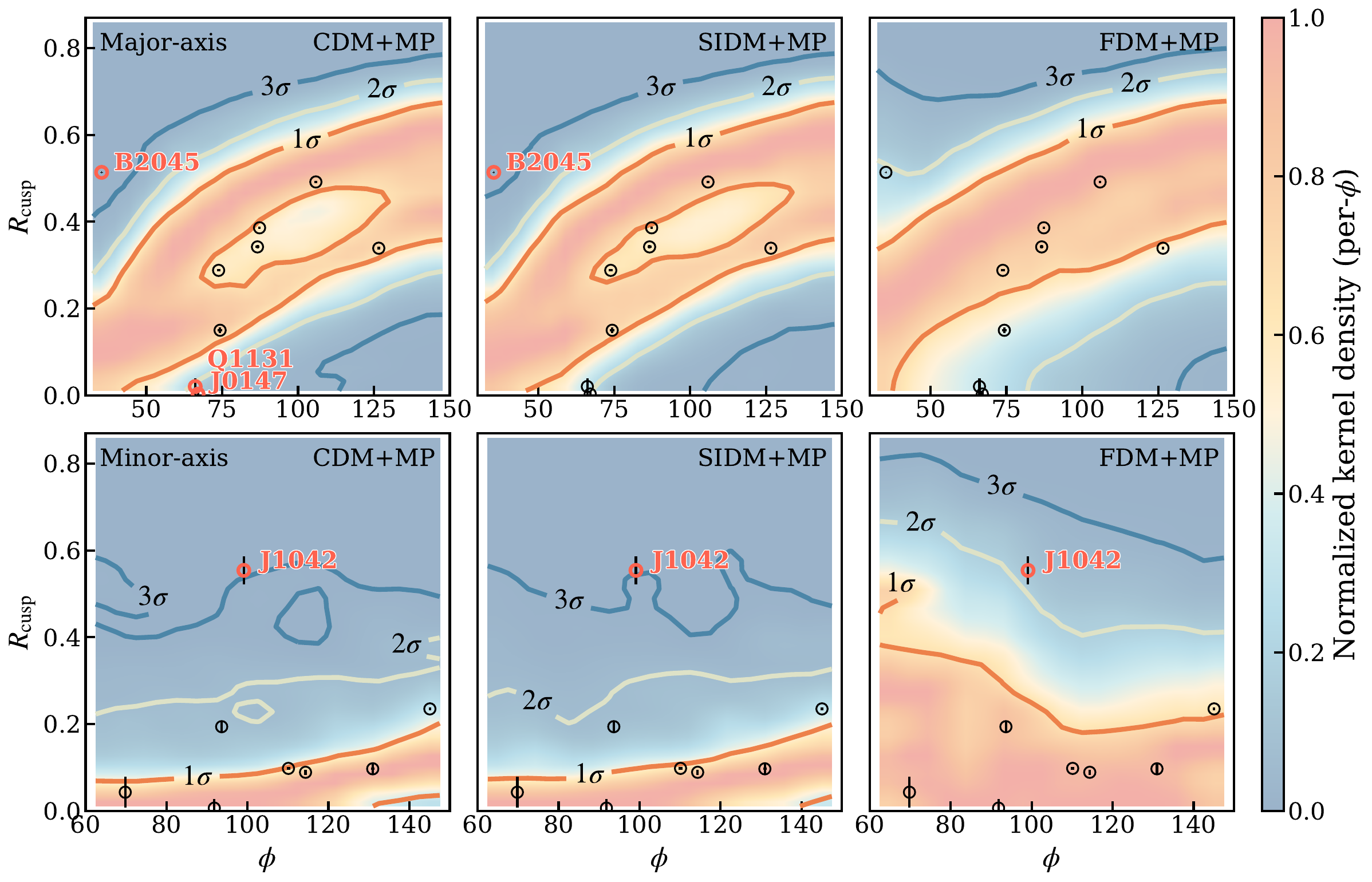}
    \caption{Same as Fig.~\ref{fig:NomonizedKDE}, but for mock lens systems generated with the SIE with external and multipole.}
     \label{fig:NomonizedKDE_MP}
\end{figure}

The KDE-normalized $R_{\rm cusp}$ distributions in Fig.~\ref{fig:NomonizedKDE_MP} are similar to those in Fig.~\ref{fig:NomonizedKDE}. 
However, the mock lenses in Fig.~\ref{fig:NomonizedKDE_MP} are generated using an SIE macromodel with external shear and higher-order multipoles.
This model is enhanced by higher-order multipoles, which introduce additional angular structure in the lens potential.

The inclusion of higher-order multipoles reduces the number of low-density regions for major-axis cusp configurations, as shown in Fig.~\ref{fig:NomonizedKDE_MP}.
Several systems that lay outside the $1\sigma$ region under the SIE with external shear model fall within the $1\sigma$ range of the CDM and SIDM predictions once multipoles are included.
This suggests that some of the tension for major-axis cusp configurations can be explained by the angular complexity of the macromodel.
However, B2045 remains a persistent anomaly in the CDM and SIDM scenarios, with its high $R_{\rm cusp}$ still located in the low-density tail, 
even after multipoles are introduced.
For minor-axis cusp configurations, intermediate-$R_{\rm cusp}$ systems, such as J0659,
lie within the $2\sigma$ density region in the CDM and SIDM scenarios,
while remaining well within the $1\sigma$ density band in the FDM scenario.
The system J1042 stands out as the only robust minor-axis cusp anomaly.
It is incompatible with both CDM and SIDM at the $>3\sigma$ level,
while remaining consistent with FDM within $3\sigma$.

In Fig.~\ref{fig:NomonizedKDE_MP}, for major-axis cusp configurations, the KDE density develops a clear valley-like structure when multipoles are included. Its upper region forms an arched high-density ridge and its lower part remains close to the original near-linear locus predicted by the pure SIE model.
Lenses with large ellipticity or higher-order multipoles occupy the upper region of the valley, 
where $R_{\rm cusp}$ increases at fixed $\phi$ (See \EDFig{fig: selection_bias} for details). 
This reflects global distortions in the critical curve due to the added angular complexity in the lens potential. 
In contrast, systems with weak multipoles maintain a near-linear $R_{\rm cusp}$–$\phi$ relationship, 
similar to the one observed in systems without multipoles. 
This suggests that multipoles mainly affect major-axis cusps at large $\phi$ ($\phi > 70^\circ$) when the lens shows significant angular structure. 
For $\phi < 70^\circ$, their impact is minimal, and fewer than 1\% of CDM or SIDM realizations occupy the region near B2045. 
Ref.~\cite{Congdon:2005iw} showed that explaining exceptionally large $R_{\rm cusp}$ for B2045 would require multipoles as high as $m \sim 9$ to 17, which cannot be produced by realistic baryonic structures. 
Such extreme multipole power is unfeasible in CDM or SIDM halos, and only FDM offers a plausible mechanism for generating perturbations of this magnitude. 

For the minor-axis cusp plots, adding multipoles does not change the qualitative separation in the smooth case (see \EDFig{fig:NomonizedKDE_MP_Smooth} for details). 
Once substructure is included, multipoles significantly broaden the CDM and SIDM distributions in the $2\sigma$–$3\sigma$ bands, while the $1\sigma$ band is only weakly affected.
This trend is likely due to a selection bias, as systems classified as major-axis cusp configurations tend to have higher ellipticity or stronger higher-order multipoles (see \EDFig{fig: selection_bias} for further discussion of this bias). 
As the ellipticity or the strength of the $m=3$ and $m=4$ perturbations increases, the caustic cross-section associated with major-axis cusp configurations grows significantly, making such systems more likely to satisfy the criteria for cusp configurations.

%According to the Bayesian evidence, Smooth-lens models remain decisively excluded even when higher-order multipoles are allowed, demonstrating that multipole freedom cannot rescue a substructure-free interpretation of the data. Table~\ref{tab:Bey} summarizes that introducing multipole terms, which allow a more flexible mass model, substantially reduces the separation among CDM+MP, SIDM+MP, and FDM+MP. The resulting Bayes-factor contrasts drop to $\mathcal{O}(1)$ in $\log_{10}\mathrm{BF}$ but remain non-negligible, with typical Bayes factors still exceeding 10. FDM+MP remains favored. Notably, the utility of multipole freedom is strongly model dependent. CDM and SIDM both benefit from the inclusion of multipoles, with evidence increases of $\Delta\log_{10}\mathrm{BF}\sim3.5$ and $\sim1.1$, respectively. In contrast, the evidence for FDM decreases by $\sim1.6$ when multipoles are added. This behavior suggests that the potential fluctuations intrinsic to FDM already account for the observed deviations in the $R_{\mathrm{cusp}}$ relation.
%\HSY{This behavior is physically well motivated.
%In FDM, mergers that form massive subhalos are suppressed, reducing the incidence of boxy or disky distortions that would otherwise generate strong higher-order multipoles.
%Consequently, additional multipole freedom does not improve—and may even reduce—the model evidence. This supports the interpretation that FDM naturally explains the observed $R_{\mathrm{cusp}}$ deviations without invoking extra angular complexity.}

%%%%%%%%%%%%%%%%%%%%%%%%%%%%%%%%%%%%%%%%%%%%%%%%%%%
\section*{Conclusion and future prospects}
%%%%%%%%%%%%%%%%%%%%%%%%%%%%%%%%%%%%%%%%%%%%%%%%%%%
 \begin{table}
    \centering
    \small
    \caption{
%    Bayesian model comparison across dark matter and lens-structure prescriptions, quantified by $\log_{10}(BF_c/BF_r)$. Positive values favor the compared model, while negative values favor the reference model. Following the Jeffreys scale, values $\gtrsim 2$ indicate decisive evidence, $\sim 1$–$2$ very strong evidence, and $\sim 0.5$–$1$ moderate evidence. The model suite includes CDM, SIDM, and FDM substructure, with and without additional higher-order multipole perturbations (+MP). Smooth lens models without dark matter substructure are decisively disfavored ($\log_{10}(BF) < -40$)
    Bayesian model comparison among dark matter scenarios and lens-structure prescriptions, quantified by the Bayes factor ratio $\mathrm{BF}{\text{(row)}}/\mathrm{BF}{\text{(column)}}$.
    The scenarios include CDM, SIDM, and FDM, with and without higher-order multipole perturbations (MP).
    Values larger than unity indicate that the model listed in the row is favored over that in the column.
    The strength of this preference is interpreted using the Jeffreys scale applied to the Bayes-factor ratio, with values $\gtrsim 100$ corresponding to decisive evidence, $\sim 10$–$100$ to very strong evidence, and $\sim 3$–$10$ to moderate evidence.
We note that smooth lens models without substructure are overwhelmingly disfavored
($\mathrm{BF}{\rm (smooth)}/\mathrm{BF}{\rm (sub)} \lesssim 10^{-35}$) (see \EDTable{tab:Bey_smooth} for details),
where $\mathrm{BF}{\rm (sub)}$ denotes models including dark-matter substructure.
    }
    \label{tab:Bey}

    \begin{tabular}{lccccc}
    \toprule
    \multicolumn{6}{c}{
    $\mathrm{BF(row)/BF(column)}$
    } \\
    \midrule
            & CDM & SIDM & CDM+MP & SIDM+MP & FDM+MP \\
    \midrule
    SIDM        & 219          &        &        &        &        \\
    CDM+MP      & 334          & 1.5    &        &        &        \\
    SIDM+MP     & $10^{3}$     & 7.8    & 5.1    &        &        \\
    FDM+MP      & $10^{4}$     & 60     & 40     & 7.7    &        \\
    FDM         & $10^{6}$     & $10^{4}$ & $10^{4}$ & 810    & 105    \\
    \bottomrule
    \end{tabular}
\end{table}

%As upcoming facilities deliver larger samples of microlensing-free, narrow-cusp lenses, this diagnostic framework offers a direct and macromodel-independent route to interpreting flux-ratio anomalies. In particular, fourth-generation surveys such as \emph{Euclid}, \emph{Roman}, and \emph{CSST} are expected to substantially increase the number of suitable cusp systems, enabling future microlensing-free flux ratios to confirm the dark matter interpretations identified in this work.

We summarize our result of Bayesian model comparison among all scenarios in Table~\ref{tab:Bey}. 
Bayesian evidence decisively excludes smooth-lens models, even when accounting for higher-order multipoles (see \EDTable{tab:Bey_smooth}). 
This confirms that multipole freedom cannot mimic the effects of dark matter substructure. 
As summarized in Table~\ref{tab:Bey}, adding multipole terms reduces the statistical gap between CDM, SIDM, and FDM, narrowing the Bayes-factor contrasts to $\mathcal{O}(10)$ in BF. Nevertheless, FDM+MP remains the statistically preferred model,
with very strong evidence ($\Delta \mathrm{BF}\simeq 40$) relative to CDM. While individual anomalies could in principle be reproduced by a finely tuned subhalo near a cusp image, such configurations require extreme geometric alignments. These possibilities are fully sampled in our Monte Carlo realizations and are found to be exceedingly rare, particularly when multiple anomalous systems are considered simultaneously.

The impact of multipole freedom is clearly model dependent. While it increases the statistical evidence for CDM ($\Delta \mathrm{BF}\simeq 334$) and SIDM ($\simeq 7.8$), it disfavors the inclusion of multipoles in the FDM case. In particular, the simpler FDM model is preferred over FDM+MP by a Bayes factor of $\sim 105$.
This is physically consistent: the suppression of massive subhalos in FDM reduces the boxy or disky distortions that typically require higher-order multipoles. 
Instead, the intrinsic potential fluctuations in FDM naturally explain the observed $R_{\text{cusp}}$ deviations. 
The fact that additional macromodel complexity does not improve the FDM fit therefore indicates that FDM provides a more direct and natural explanation for the data, while also highlighting the need for caution when introducing extra macromodel freedom.

By separating major- and minor-axis cusp configurations, we show that strong discrimination arises only for minor- and narrow major-axis cusps, while systems outside this regime are largely insensitive to dark-matter substructure (see Supplementary Information for details).
Looking ahead, upcoming surveys from \emph{Euclid}, \emph{Roman}, and \emph{CSST} will deliver substantially larger samples of quadruply imaged quasars with minor- and narrow major-axis cusps.
This study offers a robust, statistically driven framework for interpreting flux-ratio anomalies and for further testing dark matter interpretations.

\appendix

\section*{Methods}
\label{sec:method}
% --- Extended Data numbering + captions ---
\setcounter{figure}{0}
\setcounter{table}{0}

\renewcommand{\figurename}{Extended Data Fig.}%
\renewcommand{\tablename}{Extended Data Table}%

% 显示编号：Extended Data Fig. 1 / Extended Data Table 1
\renewcommand{\thefigure}{\arabic{figure}}
\renewcommand{\thetable}{\arabic{table}}

% 关键：避免与正文 Figure 1/Table 1 的超链接锚点冲突
\makeatletter
\renewcommand{\theHfigure}{ED.\arabic{figure}}
\renewcommand{\theHtable}{ED.\arabic{table}}
\makeatother
\begin{figure}
    \centering
    \includegraphics[width=\linewidth]{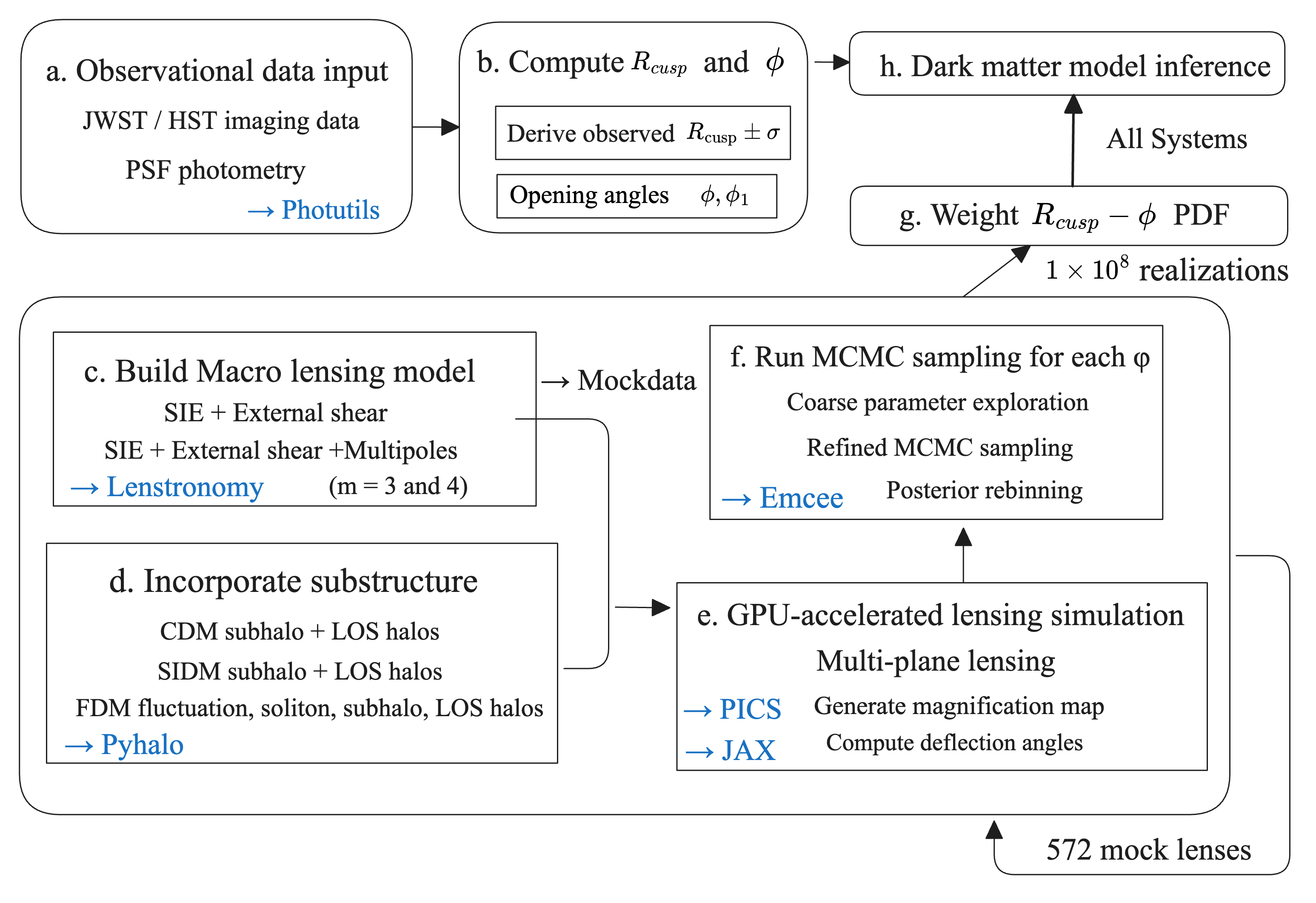}
    \caption{Workflow used in this study. (a) Observational data provide the image positions, flux ratios, and the geometric opening angle $\phi$.
(b) From these quantities, we compute the observed $R_{\rm cusp}$, which defines the constraints used to evaluate different dark matter scenarios.
(c) We construct mock macrolens models.
(d) We then add small-scale structure corresponding to CDM, SIDM, and FDM scenarios onto the smooth lens potentials.
(e) Using our \texttt{JAX}-accelerated GPU implementation of the multi-plane lensing framework, we generate the lensing realizations.
(f) A two-stage MCMC procedure samples source-plane configurations consistent with the target geometry, yielding weighted realizations of $R_{\rm cusp}$.
(g) For each dark matter scenario, we accumulate more than $10^{7}$ realizations to construct statistical predictions in the $(R_{\rm cusp},\phi)$ plane.
(h) Comparison with the observational measurements through Bayesian marginalisation then yields the relative support for competing dark matter microphysics.
    }
    
    \label{fig:Flowchart}
\end{figure}

In this section we provide the technical details of our framework. The full modeling pipeline is summarized in \EDFig{fig:Flowchart}. The framework consists of five components (observational inputs, construction of the main-lens mock data, generation of small-scale structure, multi-plane image perturbation calculations, and posterior inference), connected through a strictly forward-modeling workflow that ensures full comparability across dark matter scenarios. 

Specifically, in step (a), we extract image positions, magnifications, and the geometric opening angle $\phi$ from the observations, and compute the corresponding observed $R_{\rm cusp}$ (b). These quantities provide the geometric constraints for the Monte Carlo simulations. We then construct the macromodel mock data (c), including pure SIE and external-shear models as well as those with higher-order multipoles. On this basis (d), we generate the small-scale structure appropriate for each dark matter scenario, incorporating subhalos, LOS halos, and the interference wave field in FDM. In step (e), all lens components are embedded in a multi-plane lensing framework. We then solve the recursive lens equation using \texttt{JAX} with GPU acceleration, yielding the composite deflection field and magnification map on the image plane. To adequately sample sources near the cusp caustic, we employ a two-stage Markov Chain Monte Carlo (MCMC) scheme (f).
This ensures that each sampled source position satisfies the desired cusp geometric configuration ($\phi$, $\phi_1$, $\phi_2$).

This procedure is repeated for all mock realizations, yielding 322 (pure SIE with external shear) and 247 (SIE with external shear and multipole) sets of simulations. After completing the MCMC sampling, we normalize and weight the resulting $R_{\rm cusp}$ distributions, aggregating more than $10^7$ paired samples for each dark matter scenario (g). This provides statistical predictions for the distribution of $(R_{\rm cusp}, \phi)$ across different dark matter scenarios. Finally (h), we compare the probability density of the observed systems with the model predictions via Bayesian Decision and computation of the Bayes factor.
This yields the relative support for each dark matter scenario and thereby constrains the underlying microphysics.

This pipeline maps observations directly into the theoretical response space with minimal modeling assumptions, ensuring full consistency between geometry, lensing perturbations, and small-scale structure. It also provides a rapid, macromodel-independent diagnostic for interpreting future cusp lenses without requiring detailed per-system modeling. In the following sections, we describe each step of the procedure in detail.

\subsection{Observational Sample Details}
\label{Mocksystem}

\begin{figure}
    \centering
    \includegraphics[width=\linewidth]{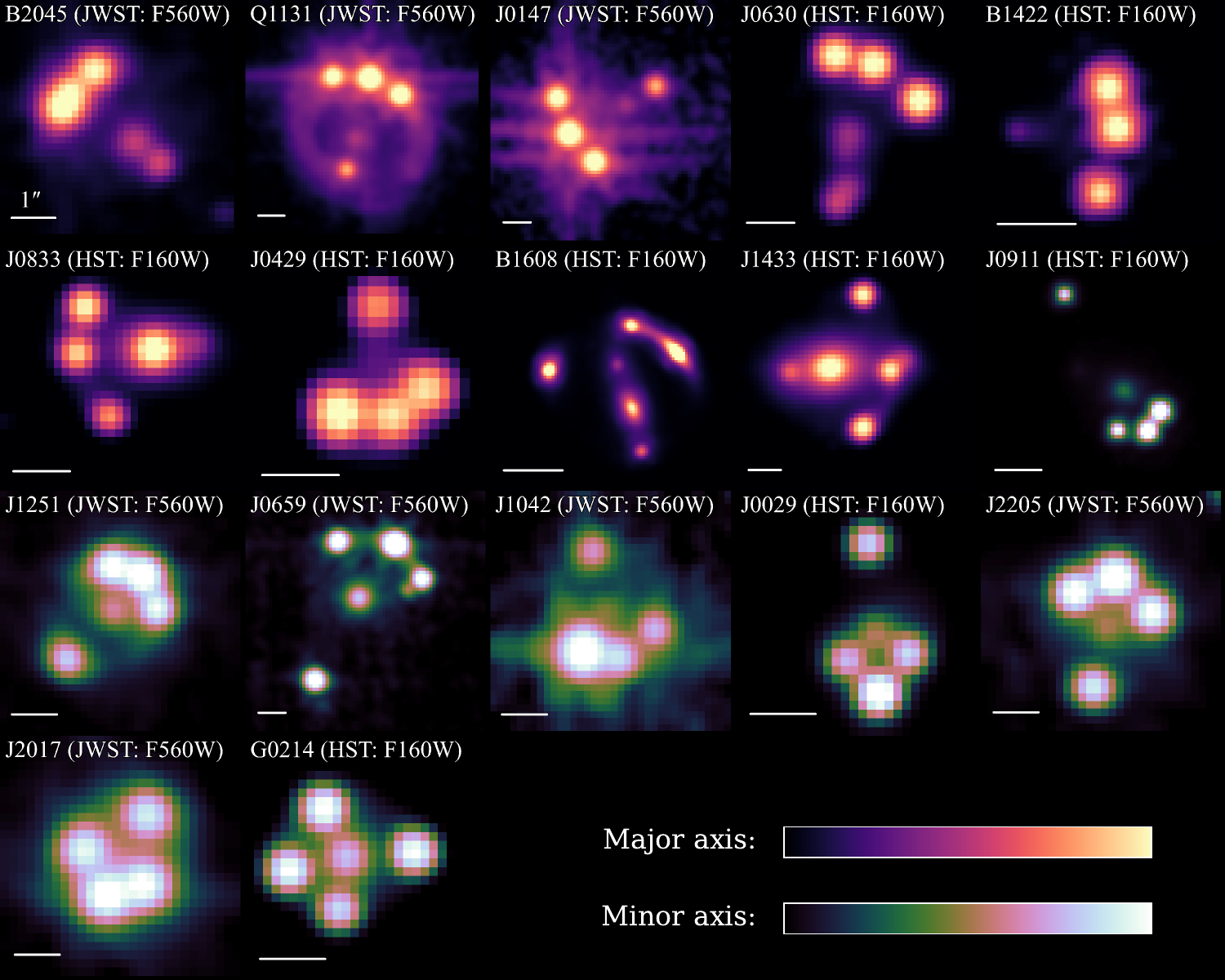}
    \caption{Background-subtracted gravitational lensing images of the selected cusp lenses. 
    The \emph{JWST} systems are taken from the GO-2046 program~\cite{Nierenberg:2023tvi,Keeley:2024brx,Keeley:2025oig,Gilman:2025fhy}, 
    while the \emph{HST} images use archival near-infrared data (see \EDTable{tab:cusp_sample_obs} for details). 
    In both cases, the wavebands adopted for the photometric measurements probe emission regions that are effectively microlensing-free, although the images shown here are for illustration only.
    Lenses are grouped into major-axis (upper two rows) and minor-axis (lower two rows) cusps,
    ordered by increasing opening angle $\phi$.
    Systems such as B2045 and J1042 exhibit pronounced flux-ratio anomalies and are of particular interest in this work.}
    \label{fig:obimage}
\end{figure}

We select cusp-configured systems from the full population of known quadruply imaged quasars using purely geometric criteria (see \EDFig{fig:Major_Minor}), without imposing any bias related to the flux ratios. As listed in \EDTable{tab:cusp_sample_obs},  this selection yields a sample of 17 cusp lenses with opening angles $\phi$ spanning $30^\circ$–$150^\circ$, where $\phi$ is defined as the maximum angle subtended by the cusp images around the lens center. Gravitational lensing images of the full sample, ordered by increasing opening angle $\phi$, are shown in \EDFig{fig:obimage}. The sample includes well-known flux-ratio anomalies such as B2045+265 and B1422+231, as well as several systems that have been less explored in previous statistical studies. This dataset represents one of the most complete samples of cusp-configured strong lenses currently available for dark matter tests.

Among the 17 cusp-configured systems in our sample, 7 have mid-infrared observations from the \textit{GO-2046 JWST} program~\cite{Nierenberg:2023tvi,Keeley:2024brx,Keeley:2025oig,Gilman:2025fhy}.
As demonstrated by ref.\cite{Keeley:2025oig}, JWST/MIRI observations at 5–21~\textmu m probe warm-dust emission regions that are insensitive to stellar microlensing, yielding microlensing-free flux-ratio measurements. We therefore adopt the warm-dust flux ratios reported in ref.\cite{Keeley:2025oig} for these 7 systems.

Both radio and narrow-line emission regions are insensitive to stellar microlensing owing to their large physical sizes~\cite{Nierenberg:2019pdj}. For the remaining 10 systems, 4 have radio or narrow-line measurements, and the corresponding flux ratios are adopted.
For the 6 systems lacking such measurements, we adopt the longest-wavelength available imaging for photometry, typically the \textit{HST} F160W band.

\begin{table}
    \centering
    \caption{Observed properties of the cusp-configuration quasar lenses used in this study. 
The table lists the system name, the image opening angle $\phi$, $R_{\rm cusp}$,
the photometric band in which $R_{\rm cusp}$ is measured, and notes on whether the lens galaxy shows obvious visible structure. A dash (-) indicates that no clear evidence for a merger is reported. For B2045, earlier claims of a merger have been revised~\cite{McKean:2006yt}, as the object previously identified as a companion galaxy is now shown to be a foreground star~\cite{Hsueh:2019ynk}.}
    \begin{tabular}{lcccc}
    \hline
    \textbf{System} 
    & $\boldsymbol{\phi}$ (deg) 
    & $\boldsymbol{R_{\rm cusp}}$ 
    & \textbf{Photometric band}
    & \textbf{Note} \\
    \hline
    \multicolumn{5}{c}{\textbf{Major-axis systems}} \\
    \hline
    B2045        & 35.30$\pm$0.04  & 0.5137$\pm$0.0003 & \textit{VLA} 15GHz~\cite{Lentz_very_2024} & - \\
    Q1131        & 66.12$\pm$0.05           & 0.0204$\pm$0.0184 & \textit{JWST} Warm~\cite{Keeley:2025oig} & - \\
    J0147        & 67.07$\pm$0.11           & 0.00311$\pm$0.0059 & \textit{JWST} Warm~\cite{Keeley:2025oig} & - \\
    J0630        & 73.83$\pm$0.60~\cite{slx173}           & 0.288$\pm$0.001   & \textit{HST} F160W~\cite{DES:2022dvw} &  Visible structure\cite{DES:2018qoi} \\
    B1422        & 74.32$\pm$0.76  & 0.150$\pm$0.006   & Narrow line $\left[\mathrm{O\uppercase\expandafter{\romannumeral3}}\right]$~\cite{Nierenberg:2014cga} & - \\
    J0833        & 86.63$\pm$0.62 & 0.342$\pm$0.004   & \textit{HST} F160W & - \\
    J0429        & 87.32$\pm$0.21           & 0.386$\pm$0.002   & \textit{HST} F160W & Visible structure~\cite{2022MNRAS509738D} \\
    B1608        & 105.87$\pm$0.43~\cite{Koopmans:2003ha} & 0.4916$\pm$0.0006 & \textit{VLA} 8.5 GHz~\cite{Fassnacht:1999re} & Visible structure~\cite{Fassnacht:1999re} \\
    J1433        & 126.59$\pm$0.076          & 0.339$\pm$0.003   & \textit{HST} F160W~\cite{DES:2022dvw} & - \\
    \hline
    \multicolumn{5}{c}{\textbf{Minor-axis systems}} \\
    \hline
    J0911        & 69.80$\pm$0.019           & 0.0431$\pm$0.0360 & Narrow line $\left[\mathrm{Ne\uppercase\expandafter{\romannumeral3}}\right]$~\cite{Nierenberg:2019pdj} & Visible structure~\cite{Nierenberg:2019pdj} \\
    J1251        & 91.81$\pm$0.15           & $0.00643^{+0.021}_{-0.018}$ & \textit{JWST} Warm~\cite{Keeley:2025oig} & - \\
    J0659        & 93.63$\pm$0.04           & 0.194$\pm$0.011   & \textit{JWST} Warm~\cite{Keeley:2025oig} & Visible structure~\cite{Gilman:2025fhy} \\
    J1042        & 99.10$\pm$0.10           & 0.554$\pm$0.032   & \textit{JWST} Warm~\cite{Keeley:2025oig} & Visible structure~\cite{2018arXiv180705434G} \\
    J0029        & 110.12$\pm$0.45          & 0.098$\pm$0.003   & \textit{HST} F160W~\cite{DES:2022dvw} & - \\
    J2205        & 114.35$\pm$0.28           & 0.089$\pm$0.006   & \textit{JWST} Warm~\cite{Keeley:2025oig} & - \\
    J2017        & 130.95$\pm$0.41          & 0.097$\pm$0.014   & \textit{JWST} Warm~\cite{Keeley:2025oig} & Spiral galaxy~\cite{Keeley:2025oig} \\
    G0214        & 145.07$\pm$0.15          & 0.235$\pm$ 0.003  & \textit{HST} F160W~\cite{DES:2022dvw} & - \\
    \hline
    \end{tabular}
    \label{tab:cusp_sample_obs}
\end{table}

\begin{table}
    \centering
    \caption{
    Redshifts and macromodel parameters of the cusp-configuration lens sample.
    An asterisk (*) indicates systems for which no reliable redshift measurement is currently available. 
    In these cases, a fiducial redshift $z_{l} = 0.5$ is adopted solely for the purpose of estimating the macromodel parameters of the main lens, including the Einstein radius $\theta_E$, the axis ratio $q$, and the external shear $\gamma$.
    }
    \begin{tabular}{lccccc}
    \hline
    \textbf{System}
    & $\boldsymbol{z_l}$ 
    & $\boldsymbol{z_s}$ 
    & $\boldsymbol{\theta_E}$ 
    & $\boldsymbol{q}$ 
    & $\boldsymbol{\gamma}$ \\
    \hline
    \multicolumn{6}{c}{\textbf{Major-axis systems}} \\
    \hline
    B2045  & 0.8673~\cite{Fassnacht:1998ay}
           & 1.28 
           & 1.91~\cite{Keeley:2025oig}
           & 0.17
           & 0.312 \\
    Q1131  & 0.295~\cite{Sluse:2003iy}
           & 0.658 
           & 1.51~\cite{Gilman:2025fhy}
           & 0.80
           & 0.11 \\
    J0147  & 0.678~\cite{Goicoechea_2019} 
           & 2.377~\cite{Rubin_2018} 
           & 1.85~\cite{Gilman:2025fhy}
           & 0.82
           & 0.16 \\
    J0630  & 0.5* & 3.34~\cite{slx173} & 1.02\cite{DES:2018qoi} & 0.53 & 0.14 \\
    B1422  & 0.339~\cite{Tonry:1997pc} 
           & 3.62~\cite{mnras25911P} 
           & 0.76~\cite{Kormann:1993pk} & 0.57 & 0.10 \\
    J0833  & 0.5* & 3.26~\cite{stac3721} & 1.15 & 0.65 & 0.01 \\
    J0429  & 0.5* & 3.866~\cite{desira_discovery_2021} & 0.704~\cite{2022MNRAS509738D} & 0.88 & 0.023 \\
    B1608  & 0.63~\cite{1995ApJ447L5M} & 1.39~\cite{1996ApJ460L103F} & 0.531~\cite{Koopmans:2003ha}  & 0.606 & 0.085 \\
    J1433  & 0.407~\cite{Agnello:2017bjk} & 2.737 & 1.71\cite{DES:2018qoi} & 0.51 & 0.09 \\
    \hline
    \multicolumn{6}{c}{\textbf{Minor-axis systems}} \\
    \hline
    J0911  & 0.769~\cite{Kneib:2000ty} 
           & 2.763 
           & 0.85~\cite{Gilman:2025fhy} 
           & 0.84 
           & 0.28 \\
    J1251  & 0.410~\cite{Kayo:2007xq} & 0.802 & 0.84~\cite{Gilman:2025fhy} & 0.72 & 0.09 \\
    J0659  & 0.766 ~\cite{Mozumdar:2022feb}
           & 3.083 
           & 2.0~\cite{Gilman:2025fhy} 
           & 0.86 
           & 0.11 \\
    J1042  & 0.599 & 2.517~\cite{Glikman_2023} & 0.89~\cite{Gilman:2025fhy} & 0.67 & 0.08 \\
    J0029  & 0.5* & 2.821~\cite{DES:2022dvw} & 0.769 & 0.54 & 0.252 \\
    J2205  & 0.63~\cite{Gilman:2025fhy} & 1.85 & 0.76~\cite{Gilman:2025fhy}  & 0.82 & 0.06 \\
    J2017  & 0.201~\cite{Stern_2021} & 1.72 & 0.59~\cite{Keeley:2025oig} & 0.32 & 0.103 \\
    G0214  & 0.22~\cite{Spiniello:2019dhi} & 3.229~\cite{Spiniello:2019dhi} &  0.849~\cite{DES:2022dvw}& 0.86 & 0.101 \\
    \hline
    \end{tabular}
    \label{tab:cusp_sample_model}
\end{table}

\subsection{Photometric and Geometric Measurements}

In a typical lensed-quasar system, the observed image contains light from the deflector galaxy, the lensed quasar host galaxy, and the quasar itself. Because the quasar emission region is unresolved at the angular resolution of current observations, we treat it as a point source described by the filter point-spread function (PSF). Accurate photometry therefore requires modeling and subtracting the extended light from the deflector and the lensed host galaxy, so that the quasar images can be cleanly isolated. The cusp statistic $R_{\rm cusp}$ is determined solely by image positions and flux ratios, thus it does not rely on any parametric macromodel. Our lens modeling is used only to remove the lens and host light and to isolate the quasar point sources. It is also used to extract point-source fluxes under a PSF description, but it is not used to constrain or validate the deflector mass model. Consequently, the inferred flux ratios and hence $R_{\rm cusp}$ are driven by the imaging data and the PSF and light modeling, rather than by assumptions about the macromodel.

\begin{figure}
    \centering
    \includegraphics[width=\linewidth]{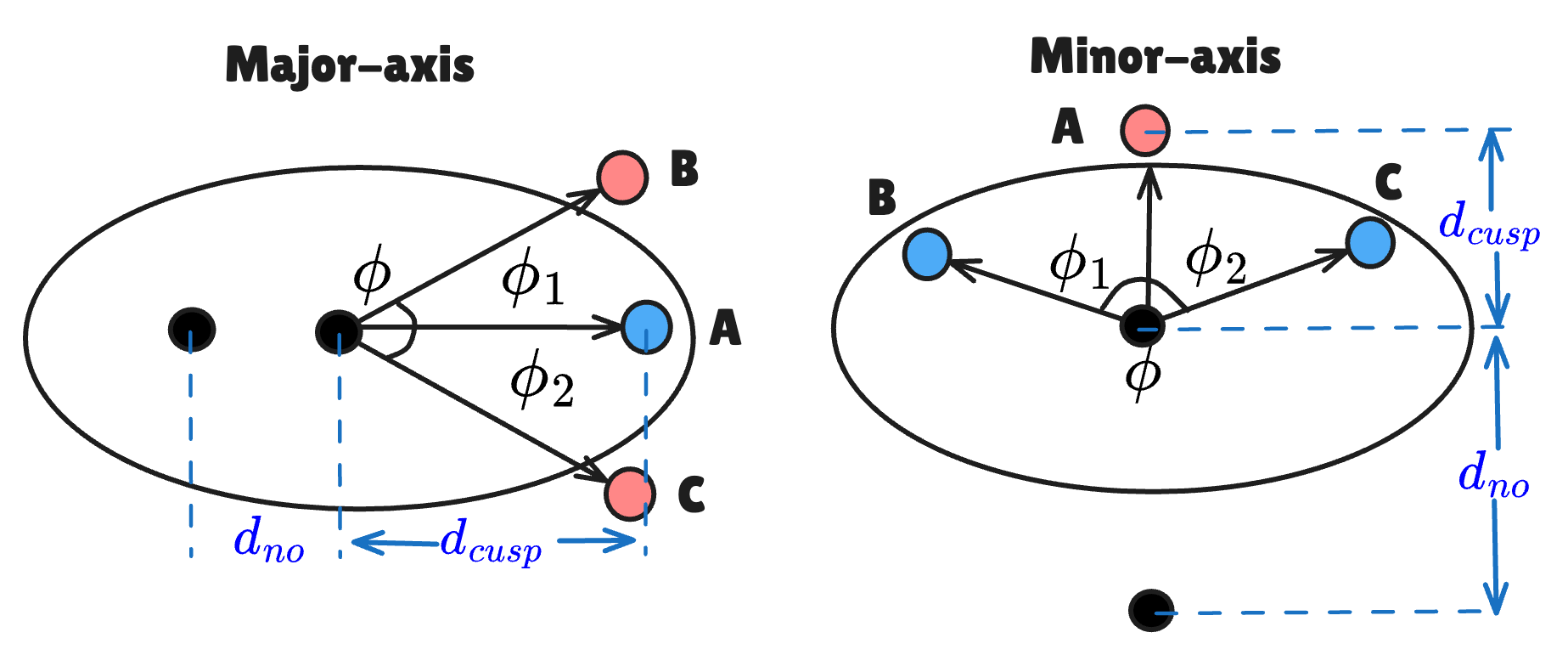}
    \caption{Schematic illustration of the geometric criterion for distinguishing major- and minor-axis cusp configurations in an elliptical lens. 
    The distances from the lens center to the non-cusp image and to the midpoint of the cusp triplet are denoted by $d_{\rm no}$ and $d_{\rm cusp}$, respectively. 
    Systems with $d_{\rm cusp} > d_{\rm no}$ are classified as major-axis cusps, while those with $d_{\rm cusp} < d_{\rm no}$ are minor-axis cusps. 
    The colours of the three cusp images indicate the sign of the magnification (red: positive parity; blue: negative parity). 
    The opening angle $\phi$ is defined by the two outer cusp images, and $\phi_{1}$ and $\phi_{2}$ denote the directions from the central image to its two companions. 
    This geometric diagnostic can be identified directly from observed lensing morphologies and provides a practical criterion for classifying cusp configurations.}
     \label{fig:Major_Minor}
\end{figure}

For a typical cusp-configured quadruply imaged quasar lens, three of the images form close to a cusp of the tangential critical curve. We define $\phi$ as the opening angle measured at the deflector center between the two outer images of this triplet, as illustrated in \EDFig{fig:Major_Minor}. Accurate astrometry for systems with point-like quasar images is readily obtained from short-wavelength observations. The PSF is compact and the image centroids are tightly constrained. If the deflector center is not clearly detected in the imaging data, we determine its position by fitting the lens-galaxy light distribution. This yields a well-defined reference center for measuring $\phi$ consistently across the sample. The uncertainty in $\phi$ is computed via standard error propagation from the positional uncertainties. For the systems in our sample, the resulting uncertainties are typically $\lesssim 1^\circ$.

As illustrated in \EDFig{fig:Major_Minor}, the cusp triplet lies close to either the projected major- or minor-axis of the deflector's mass distribution. We therefore classify each system as a major-axis cusp or a minor-axis cusp based on which axis the cusp images are nearest to.
When the source lies near the cusp on the long axis of the astroid caustic, the resulting image arrangement forms what we classify as a major-axis cusp configuration. Conversely, when the source approaches the short-axis cusp, the system is classified as a minor-axis configuration. 
In practice, we classify a system as a major-axis or minor-axis cusp by reading off the geometry in \EDFig{fig:Major_Minor}. The cusp triplet is identified as the three images that cluster on the same side of the deflector. The remaining, more isolated image is taken to be the non-cusp image. We then measure two distances from the lens center. The quantity $d_{\rm cusp}$ is defined as the distance to the midpoint of the triplet, while $d_{\rm no}$ is defined as the distance to the isolated image. If $d_{\rm cusp}>d_{\rm no}$ the system is a major-axis cusp, whereas $d_{\rm cusp}<d_{\rm no}$ indicates a minor-axis cusp. This simple geometric rule is robust for lenses with appreciable ellipticity.

For lenses that look nearly circular on the sky, it becomes harder to distinguish major-axis cusps from minor-axis cusps.
In such cases the image geometry alone may be insufficient to assign a unique classification. As indicated by \EDFig{fig: selection_bias}, the distinction between the major and minor-axis cusp branches in the $R_{\rm cusp}$–$\phi$ plane diminishes as the projected ellipticity decreases. For nearly circular systems, the two branches gradually converge. Since in the limit of an axisymmetric lens the notions of long and short axes become degenerate and the cusp relation becomes insensitive to orientation.
Where the image geometry is ambiguous, the classification is cross-checked with lens modeling. The reconstructed source position with respect to the astroid caustic directly indicates whether the source lies near a major-axis or minor-axis cusp.

\subsection{Simulation Sample Details}
\label{Sec:Simulation Sample Details}

%To interpret this observational sample, we adopt an all-sky mock lensing framework. This approach provides a macromodel-independent and scalable route to forward modeling, enabling a statistical comparison among different dark-matter scenarios. We construct a baseline set of smooth lensing mock data using an idealized Singular Isothermal Ellipsoid (SIE) profile \cite{1994A&A...284..285K}, supplemented by an external shear component, following the framework developed by~\RefCite{Dong:2024sij}.

To constrain dark matter properties using the statistics of lensing flux-ratio observations, we employ an all-sky mock lensing framework~\cite{Dong:2024sij}, including the procedures for creating lens and source populations, lensing calculations, and selecting lensing systems. 

\subsubsection*{Lens population: SIE + shear}

First, the primary lens mass is modeled with the Singular Isothermal Ellipsoid (SIE) profile \cite{1994A&A...284..285K}. Hence, the corresponding convergence $\kappa(x,y)$ is given by
\begin{equation}
\kappa(x,y)
=
\frac{\theta_{\rm E}}{2}\,
\frac{\lambda(e)}
{\sqrt{(1-e)^{-1}x^{2} + (1-e)y^{2}}},
\end{equation}
where $e$ denotes the ellipticity of the projected mass distribution, and $\lambda(e)$ is a dynamical normalization factor that accounts for the three-dimensional shape of the lens galaxy~\cite{Chae:2002uf, Oguri:2010ns}. The Einstein radius $\theta_{\rm E}$ is related to the LOS velocity dispersion $\sigma_v$ via $\theta_{\rm E}=4\pi(\sigma_v/c)^2(D_{\rm ls}/D_{\rm s})$, where $D_{\rm ls}$ and $D_{\rm s}$ denote the angular diameter distances from the lens to the source and from the observer to the source, respectively. Thus, it is sufficient to describe a primary lens with the parameters of $(x_0, y_0, \sigma_v, e, z_l)$, where $(x_0, y_0)$ denotes the lens center and $z_l$ is the lens redshift.

For creating the population of deflectors, we resample the velocity dispersion $(\sigma_v)$ and redshift $(z_l)$ following a redshift-dependent velocity dispersion distribution function (VDF): 
\begin{equation}
n(\sigma_v,z_l)=n_\ast(z_l)\left(\frac{\sigma_v}{\sigma_\ast(z_l)}\right)^a
\exp\!\left[-\left(\frac{\sigma_v}{\sigma_\ast(z_l)}\right)^b\right]\frac{1}{\sigma_v},
\end{equation}
where $n(\sigma_v,z_l)$ is the comoving number density of galaxies with velocity dispersion $\sigma_v$ at lens redshift $z_l$. 
We adopt the observationally motivated evolution from ~\RefCite{Yue:2022lcc}, with $n_\ast(z_l)=6.92\times10^{-3}(1+z_l)^{-1.18}\,{\rm Mpc}^{-3}$ and $\sigma_\ast(z_l)=172.2(1+z_l)^{0.18}\,{\rm km\,s^{-1}}$, and fixed shape parameters $a=-0.15$ and $b=2.35$. This VDF is valid up to $z_l\sim1.5$ and includes both quiescent and star-forming galaxies. We consider lenses in the range $0<z_l<2$ with $100<\sigma_v<450~{\rm km\,s^{-1}}$, corresponding to a total of $\sim8.6\times10^8$ potential lenses over the full sky, which we assume to be uniformly distributed, thus the angular positions of the deflectors are ready. The ellipticity $e$ is drawn from a truncated normal distribution with a mean of 0.3 and a standard deviation of 0.16, bounded between $e=0$ and $e=0.9$, which is consistent with observationally inferred ellipticity distributions of early-type galaxies \cite{SDSS:2007jtw}.

Besides, we involve external shears to model environmental effects, and the lensing potential is given by
\begin{equation}
\psi(x,y)=\frac{\gamma}{2}(x^{2}-y^{2})\cos2\theta_\gamma-\gamma xy\sin2\theta_\gamma,
\end{equation}
which is derived from N-body cosmological simulations combined with semi-analytic galaxy formation models~\cite{Holder:2002hq}. The shear strength $\gamma$ follows a log-normal distribution with $\langle\log_{10}\gamma\rangle=-1.301$ and dispersion $0.2$ dex, and the orientation $\theta_\gamma$ is uniformly random in $[0^\circ,360^\circ]$.

\subsubsection*{Lens population: multipoles}

SIE macromodels impose overly restrictive symmetry assumptions.
However, real lens galaxies often depart from pure ellipticity, driven by disks, asymmetric stellar light, or merger remnants, and are therefore inconsistent with SIE predictions.
To capture these effects, we construct an extended mock sample by adding higher-order multipole perturbations to the lens potential.
The $m=3$ term introduces asymmetric distortions, whereas $m=4$ produces boxy/disky isophotes, as commonly associated with merger- or disk-induced perturbations.

\begin{figure}
    \centering
    \includegraphics[width=\linewidth]{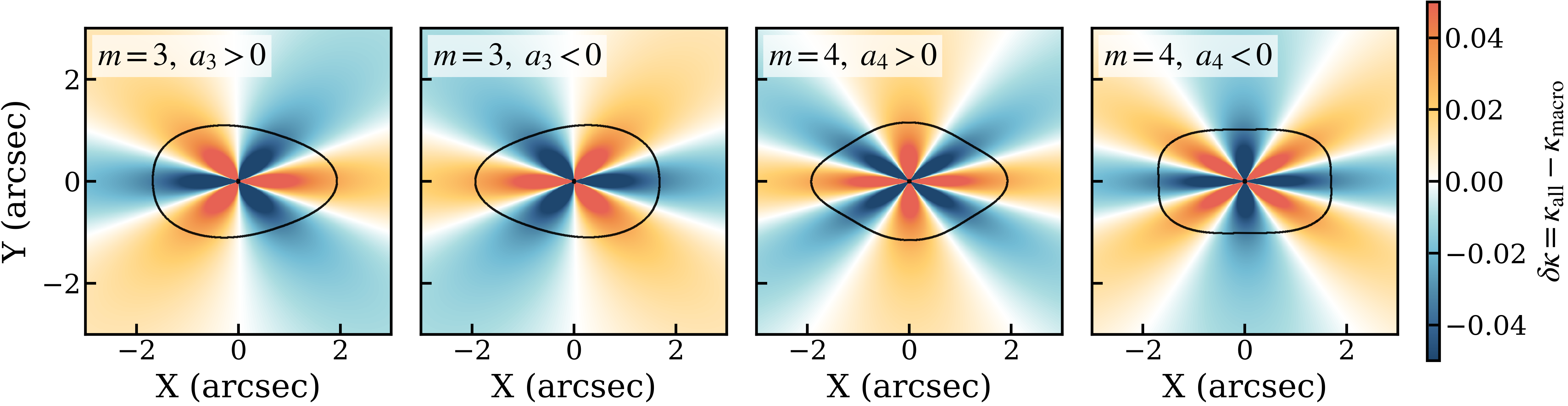}
    \caption{
        Difference in convergence induced by elliptical multipole perturbations (m=3 or 4).
        }
        \label{fig:multipoleShow}
\end{figure}

Expressly, in polar coordinates $(r,\phi)$, the radial deviation of an $m$-th order multipole is written as
\begin{equation}
\delta r(\phi)
=
a_m \cos\!\left[m(\phi-\phi_m)\right],
\end{equation}
where $\phi_m$ is the multipole orientation and $a_m$ is the perturbation amplitude. As shown in \EDFig{fig:multipoleShow}, elliptical multipoles imprint coherent, large-scale distortions in the convergence field, with $m=3$ and $m=4$ modes producing qualitatively distinct angular patterns. In particular, positive $m=4$ yields disky isophotes, whereas negative $m=4$ produces boxy isophotes.

Assuming geometric similarity between isodensity contours, the multipole amplitude scales with the size of the ellipse, $a_m\propto a$, where $a$ is the semi-major axis. We therefore characterize the multipole strength using the dimensionless ratio $a_m/a$. Following standard convention, we adopt the $\kappa=1/2$ isodensity contour as the reference, for which $a(\kappa=1/2)=\theta_{\rm E}/\sqrt{q}$. 

%To incorporate higher-order multipoles in a manner consistent with observations, we construct elliptical multipole perturbations via an elliptical coordinate transformation of the standard circular formalism \cite{Paugnat:2025bdn}. We implement these using the \texttt{MULTIPOLE\_ELL} models in \texttt{lenstronomy}.
To incorporate higher-order multipoles in a manner consistent with observations, we construct elliptical multipole perturbations via an elliptical coordinate transformation of the standard circular formalism \cite{Paugnat:2025bdn} using the \texttt{MULTIPOLE\_ELL} models in \texttt{lenstronomy}\footnote{\url{https://github.com/lenstronomy/lenstronomy}}\cite{Birrer:2018xgm, Birrer2021}.

Utilizing empirical priors from ~\RefCite{Oh:2024dlj}, derived from observed galaxy isophotes, we adopt the inferred distributions of multipole parameters describing departures from elliptical mass distributions (see \EDFig{fig:a3_left_a4q_right}). 
In these priors, the $m=3$ mode has a normalized amplitude $a_3/a$ symmetrically distributed about zero, whereas the $m=4$ mode shows systematically larger and preferentially positive $a_4/a$ in more flattened systems, consistent with disky distortions.

\subsubsection*{Source population}

The background quasar population is created by drawing sources from redshift-dependent quasar luminosity functions (QLFs) calibrated to current observations \cite{Yue:2022lcc}. The QLF is parameterized as a double power-law of the form 
\begin{equation}
\Phi(M, z_s) =
\frac{\Phi_\ast(z_s)}
{10^{0.4\,(M - M^\ast(z_s))(\alpha(z_s)+1)}
+ 10^{0.4\,(M - M^\ast(z_s))(\beta(z_s)+1)}},
\end{equation}
where $\alpha(z_s)$ and $\beta(z_s)$ are the faint-end and bright-end slopes, $M^\ast(z_s)$ is the characteristic break magnitude, and $\Phi_\ast(z_s)$ is the normalization, with their specific numerical values listed in Table~1 of~\RefCite{Yue:2022lcc}. Quasars satisfying this QLF are then distributed uniformly over the sky.

\subsubsection*{Selection criteria}

%We then ray-trace all lens configurations and apply a sequence of selection criteria to retain only systems that would be observationally accessible as quad quasar lenses. 
% We then ray-trace all QSO-deflector systems to generate candidates for lensed QSOs and apply a sequence of selection criteria to retain systems that are observationally accessible as quad-quasar lenses.
We then ray-trace all quasar–deflector systems to generate candidates for
lensed quasars and apply a sequence of selection criteria to retain systems
that are observationally accessible as quad-quasar lenses.
Specifically, a quad is classified as a cusp configuration if it meets the angular conditions $\left|\phi_1 - \phi/2\right| \le 5^\circ$ and $\left|\phi_2 - \phi/2\right| \le 5^\circ$, where $\phi$ denotes the triplet opening angle and $\phi_{1,2}$ are the angular offsets along the two characteristic directions (see \EDFig{fig:Major_Minor}). 
The observed systems considered in this work satisfy the same criteria.

In the all-sky mock catalogue, we obtain a total of $588{,}856$ strong-lensing systems.
For the SIE with shear population, $4.03\%$ ($23{,}746$ four-image systems) are quads, of which $9.97\%$ ($2{,}367$ cusp systems) satisfy our cusp-geometry criterion.
For lenses including multipoles with $m=3$ or $4$, the corresponding fractions are $4.13\%$ ($24{,}363$ four-image systems) and $10.35\%$ ($2{,}523$ cusp systems).
Thus, cusp configurations constitute $\simeq0.4\%$ of all strong lenses in either modelling family.

\begin{figure}
    \centering
    \includegraphics[width=\linewidth]{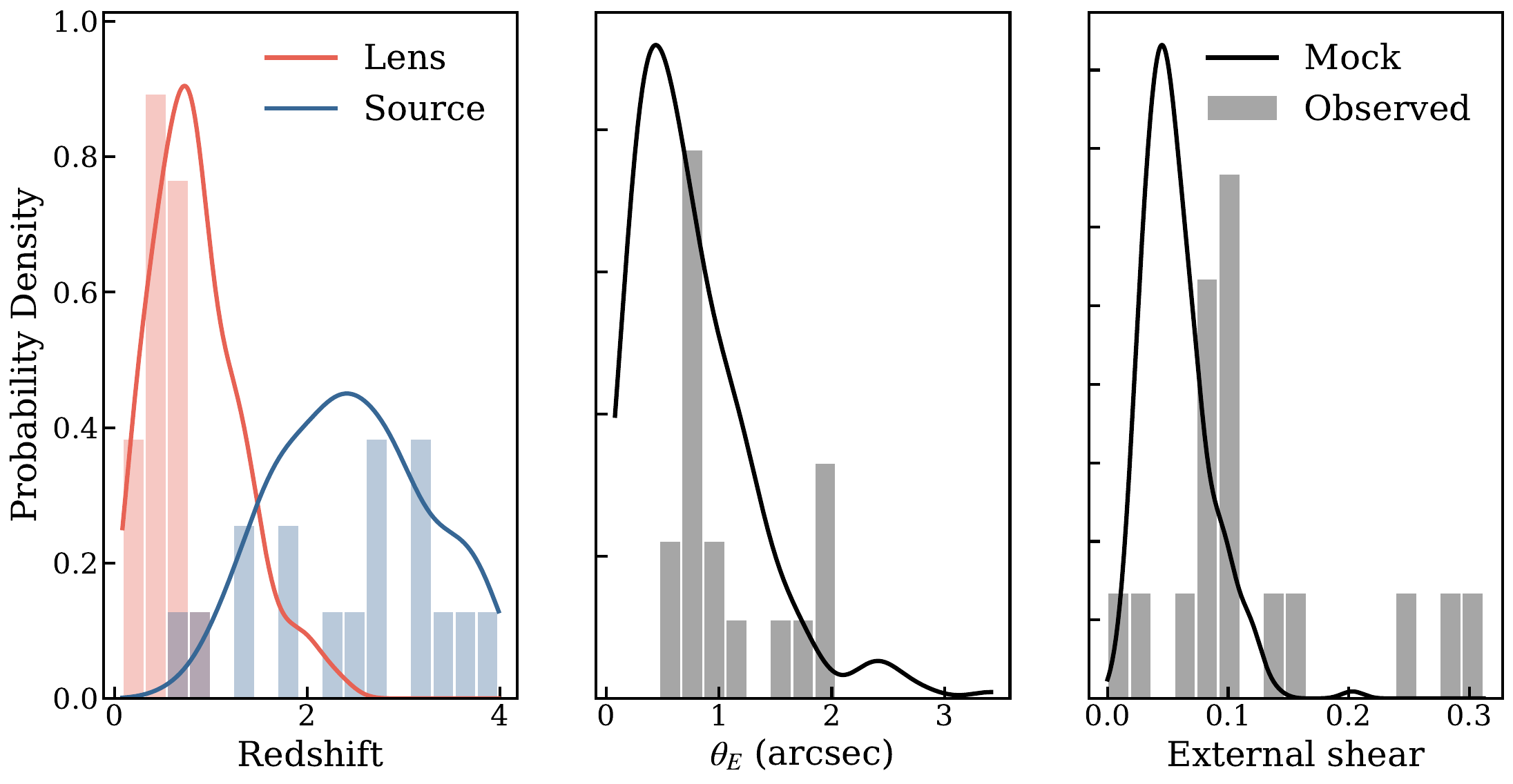}
    \caption{
        Parameter distributions of the mock sample used for the pure SIE simulations. 
        The mock distributions (lines) closely match the observed sample (histograms) in redshift, Einstein radius $\theta_{\rm E}$, and external shear, indicating that the mock catalog is representative of the data.
        }
        \label{fig:mock_sie_dist}
\end{figure}

Observability cuts are then applied to the mock sample.
Resolved four-image systems are required, with a minimum image separation of $0.07''$ (resolvability), a first-arriving image magnitude $m_{i,\mathrm{first}}<26.5$ (detectability), and a source redshift $z_{\rm s}<4$ (source visibility).
After all selections, 327 cusp systems are obtained from SIE models with external shear, and 248 from SIE models with external shear including higher-order multipoles. Distributions of lens and source redshifts, Einstein Rings, and external shears of these selected systems are shown in \EDFig{fig:mock_sie_dist}. 
\begin{figure}
    \centering
    \includegraphics[width=\linewidth]{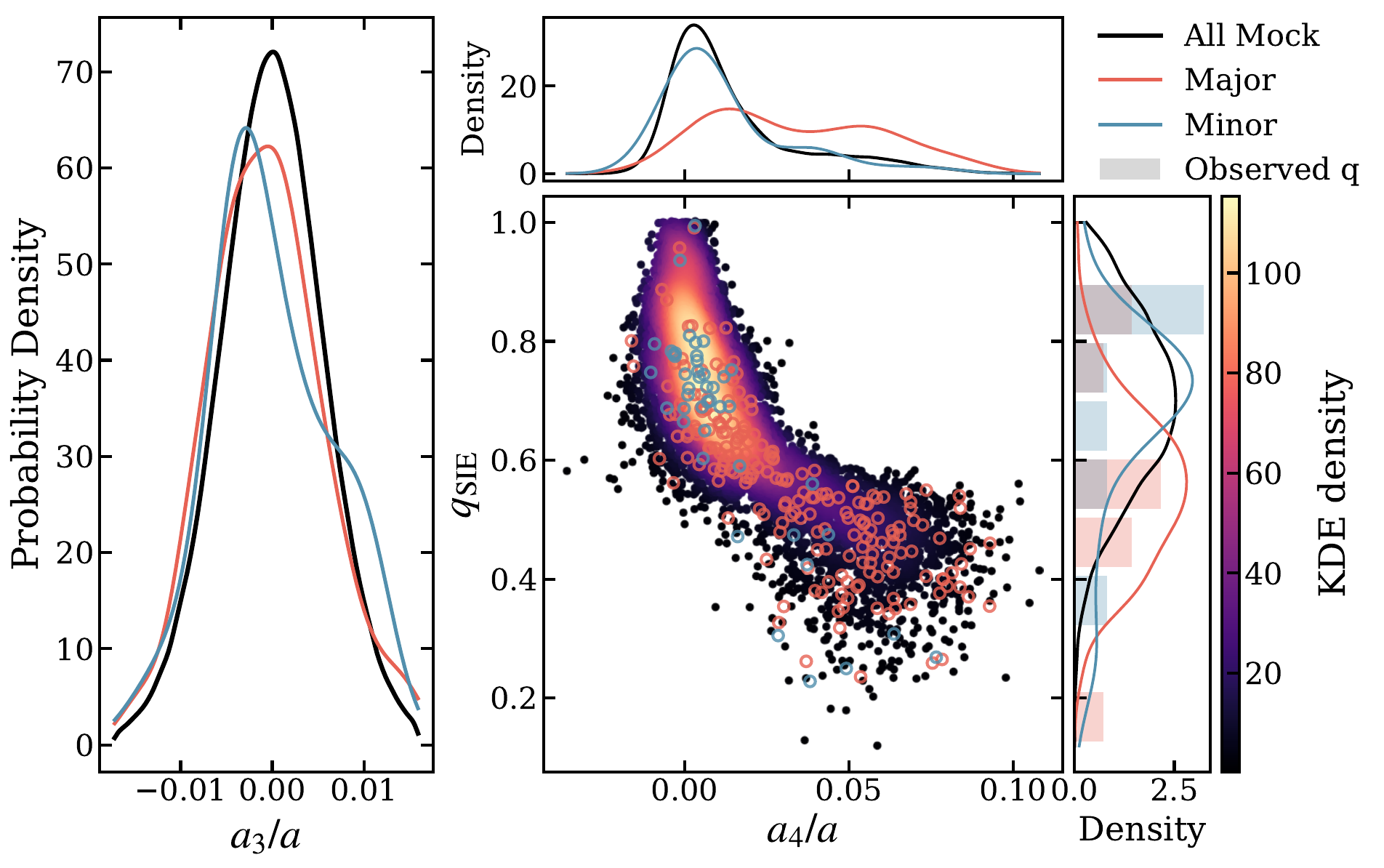}
    \caption{Selection effects on the multipole parameters in the mock lens sample. 
The left panel shows the distribution of $a_{3}/a$, for which the selected major-axis (red) and minor-axis (blue) subsamples closely match the full mock sample, indicating no significant bias in the $m=3$ multipole. 
The right-hand panels show the joint distribution of $(a_{4}/a, q_{\rm SIE})$ and the corresponding marginal distributions. 
Colormap-shaded points in the central panel represent the full mock sample, while hollow symbols mark systems passing the selection criteria. 
In contrast to $a_{3}/a$, major-axis cusp systems preferentially arise from lenses with higher ellipticity or stronger $m=4$ multipole perturbations.}
    \label{fig:a3_left_a4q_right}
\end{figure}
\begin{figure}
    \centering
    \includegraphics[width=\linewidth]{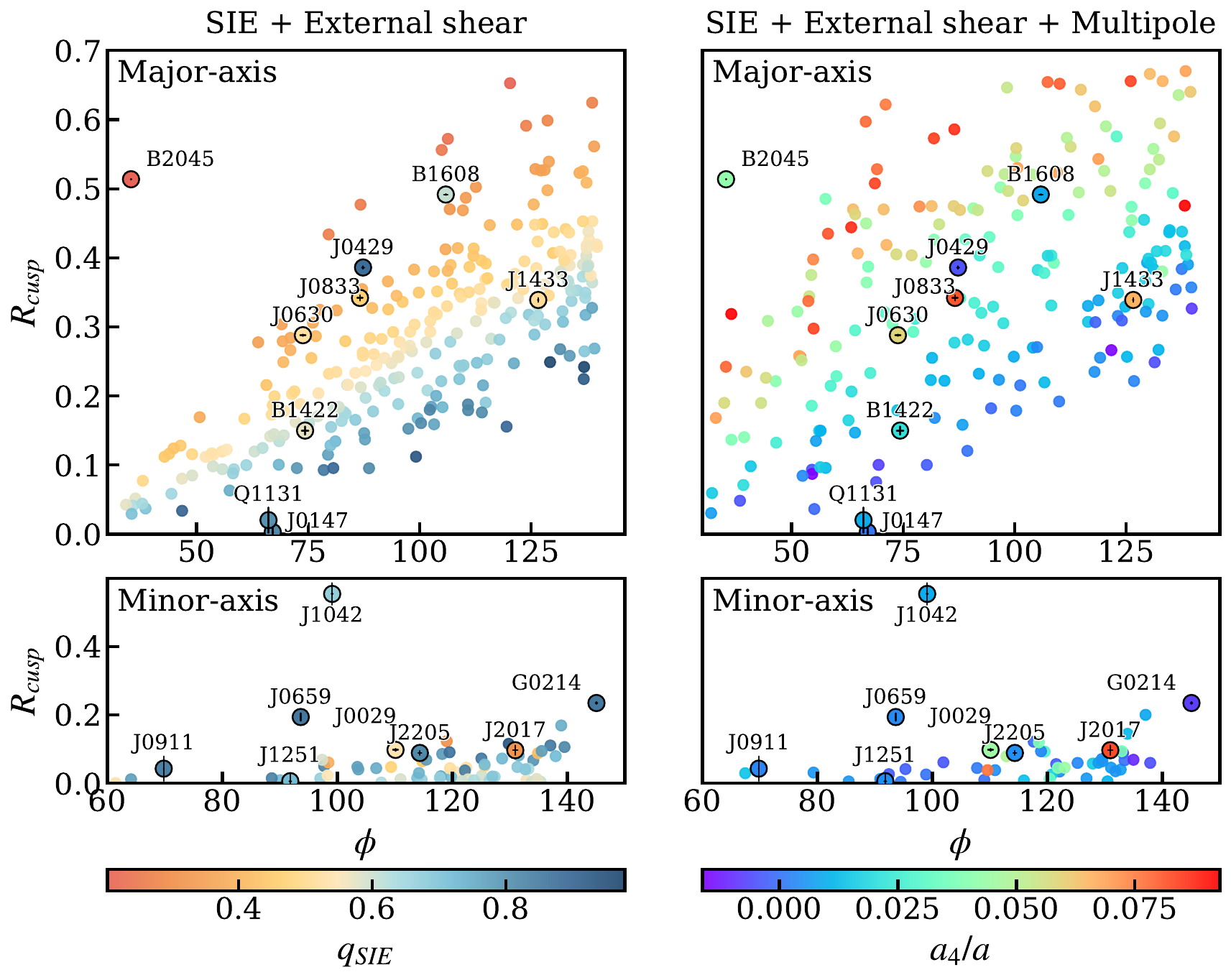}
    \caption{Comparison of the cusp relation $R_{\rm cusp}$ versus the opening angle $\phi$ for the Smooth lens. 
    Left: SIE plus external shear, showing major-axis and minor-axis cusp configurations, color-coded by the axis ratio $q_{\rm SIE}$. 
    Right: SIE models with added multipole perturbations, color-coded by $a_{4}/a$, with the values for the observed systems converted from $q_{\rm SIE}$ using the empirical prescription of ~\RefCite{Oh:2024dlj}.
    For fixed ellipticity, the $R_{\rm cusp}$–$\phi$ relation collapses into a one-dimensional locus, reflecting the constraints imposed by a smooth mass distribution. }
    \label{fig: selection_bias}
\end{figure}

The joint distributions of axis ratio and $a_{4}/a$ reveal a clear selection bias (\EDFig{fig:a3_left_a4q_right}).
This bias arises because the caustic cross section for major-axis cusp configurations increases rapidly with ellipticity and with the presence of low-order multipole structure. In other words, the same angular complexity that boosts the probability of producing a major-axis cusp can also shift its location in the $R_{\rm cusp}$ plane, making multipoles comparatively effective for this subset.

As shown in \EDFig{fig: selection_bias}, the scatter diagram of $R_{\rm cusp}$–$\phi$ collapses into an extremely narrow, nearly one-dimensional curve, illustrating the tight constraints imposed by a smooth mass distribution. 
The slope of this locus is strongly correlated with the ellipticity and quadrupole strength of the primary lens. By contrast, the observed lenses occupy a markedly broader region of the plane. 
For intermediate opening angles, part of the excess scatter in major-axis systems can be reduced by allowing multipole perturbations ($m=3,4$). This behavior is expected because major-axis cusp configurations are preferentially selected from lenses with higher ellipticity as well as stronger $m=3$ and $m=4$ amplitudes (See \EDFig{fig:a3_left_a4q_right}). 

Even with the inclusion of multipoles, two discrepancies remain. Minor-axis systems lie systematically outside the smooth-model envelope across the full parameter range, indicating that multipoles do not provide a comparable lever arm for minor-axis cusps. In addition, the widely studied major-axis lens B2045+265 shows an extreme offset that any smooth mass model cannot reproduce. These pronounced departures define a set of systems that are prime targets for detailed modeling under alternative dark matter scenarios.

\subsection{Dark Matter Substructure}
\label{Sec:Dark Matter Substructure}
Having established the macromodel of the primary lens, we next incorporate small-scale perturbations associated with dark-matter substructure. All realizations are generated using \texttt{pyHalo}, which implements semi-analytic prescriptions based on \textit{Galacticus}~\cite{Benson:2010kx} to model subhalo tidal evolution, including tidal stripping, tidal heating, and dynamical friction~\cite{Du:2025xqi}.

For CDM and SIDM, the lens model includes subhalos associated with the main deflector as well as line-of-sight (LOS) halos. In the CDM case, subhalos are distributed within the host halo following a projected NFW profile. SIDM realizations are constructed by mapping each CDM halo to its self-interacting counterpart using the fluid-based PSIDM model of Hou et al.~(2025)~\cite{Hou:2025gmv}.

For FDM, small-scale subhalos are largely suppressed. Instead, the halo is modeled with a central solitonic core and granule-like interference patterns within the same \texttt{pyHalo} framework. Our FDM implementation extends \texttt{pyHalo} by sampling density fluctuations at multiple radii, with granularity strongest near the halo center and decreasing outward. Across all dark-matter scenarios, substructure lensing is implemented through analytic calculations of the deflection angles. In the following sections, we present the detailed implementations of these three dark-matter scenarios.

\subsubsection*{CDM}
We model the internal structure of all halos using an NFW density profile,
\begin{equation}
\rho(r) = \frac{\rho_{s,0}}{(r/r_{s,0})\left(1+r/r_{s,0}\right)^2}.
\end{equation}
For CDM, all subhalos and LOS halos are assumed to follow this profile.

The subhalo mass function adopted in \texttt{pyHalo} is given by
\begin{gather}
    \frac{d^{2}N_{\rm sub}}{dm_{\rm sub}\, dA}
    = \frac{\Sigma_{\rm sub}}{m_0}
    \left(\frac{m_{\rm sub}}{m_0}\right)^{\alpha}
    \mathcal{F}(M_{\rm halo}, z), \\
    \log_{10}(\mathcal{F}) = k_1 \log_{10}\left(\frac{M_{\text{halo}}}{10^{13} M_\odot}\right) 
    + k_2 \log_{10}(z + 0.5),
\end{gather}
where \( \Sigma_{\text{sub}} = 0.025 \), \( \alpha = -1.9 \), and \( m_0 = 10^8\,M_\odot \). 
The scaling function \( \mathcal{F}(M_{\text{halo}}, z) \) accounts for the redshift and host-mass dependence of the projected number density, with \( k_1 = 0.55 \) and \( k_2 = 0.37 \)~\cite{Gannon:2025nhr}. 
Here \( M_{\text{halo}} \equiv M_{200} \) denotes the host halo mass, and \( m_{\rm sub} \) is the subhalo mass.

For LOS halos, \texttt{pyHalo} adopts the Sheth--Tormen halo mass function~\cite{Sheth:1999su}, multiplied by an overall rescaling factor $\delta_{\text{los}}$, which is assigned a uniform prior in the range [0.8, 1.2] to account for theoretical uncertainties in the mass-function normalization~\cite{Despali:2015yla}. 
In addition, a two-halo term $\xi_{\text{2halo}}(M_{\text{halo}}, z)$ is included to capture the contribution from correlated large-scale structure surrounding the main lens halo. 
The resulting LOS halo mass function is
\begin{gather}
\frac{d^2 N_{\text{los}}}{dm_{\rm los}\, dV} 
= \delta_{\text{los}}\left( 1 + \xi_{\text{2halo}}(M_{\text{halo}}, z) \right) 
\frac{d^2 N}{dm_{\rm los}\, dV} \Big|_{\text{ShethTormen}} .
\end{gather}

Halo concentrations are computed using the updated mass--concentration relation of \cite{Diemer:2018vmz}, which provides an analytic redshift dependence and remains well calibrated at the low-mass end. 
This prescription is implemented in \texttt{pyHalo} and propagated into the subhalo tidal-evolution module, ensuring that the internal structure of each halo is treated self-consistently throughout its evolution. 
We define the concentration as $c \equiv r_{200}/r_{s,0}$.
Given a halo mass $M_{200}$ and concentration $c$, the corresponding NFW profile is specified by the scale radius $r_{s,0} = r_{200}/c$ and scale density $\rho_{s,0} = M_{200}\,[4\pi r_{s,0}^{3}(\ln(1+c)-c/(1+c))]^{-1}$.

In this work, the parameters $(M_{200},c)$ are assigned by matching each mock SIE lens to an equivalent spherical mass profile extrapolated to $r_{200}$. Specifically, we define $r_{200}$ as the radius at which the mean enclosed density of the SIE profile equals $200$ times the critical density at the lens redshift, $\rho_{\rm SIE}(r_{200}) = 200\,\rho_{\rm crit}(z_l)$, and determine $r_{200}$ numerically. The associated halo mass is then estimated by integrating the SIE density profile, $M_{200}^{\rm (SIE)} = 4\pi \int_{0}^{r_{200}} r^2\,\rho_{\rm SIE}(r)\,dr$. This procedure provides a approximate mapping from the inner lensing mass model to a halo-scale mass, which we use to normalize the host NFW halo. 

\subsubsection*{SIDM}

We consider a velocity-dependent SIDM model which features a $v^{-4}$ dependence such that halos of mass lower than $10^8~\rm M_{\odot}$ are primarily core collapsing and more massive halos, particularly the $10^{11}~\rm M_{\odot}$ ones of relevance in~\RefCite{Zhang:2025bju}, are core forming. The cross section per mass is parameterized following the Rutherford scattering cross section~\cite{Feng:2009hw,Ibe:2009mk}
\begin{gather}
    \frac{d\sigma}{d\cos\theta}
    = \frac{\sigma_0\, w^4}{2\left[w^2 + v^2 \sin^2(\theta/2)\right]^2},
\end{gather}
where $\sigma_0/m = 8000\,\mathrm{cm^2\,g^{-1}}$ and $w = 6\,\mathrm{km\,s^{-1}}$. 

\begin{figure}
    \centering
    \includegraphics[width=0.472\linewidth]{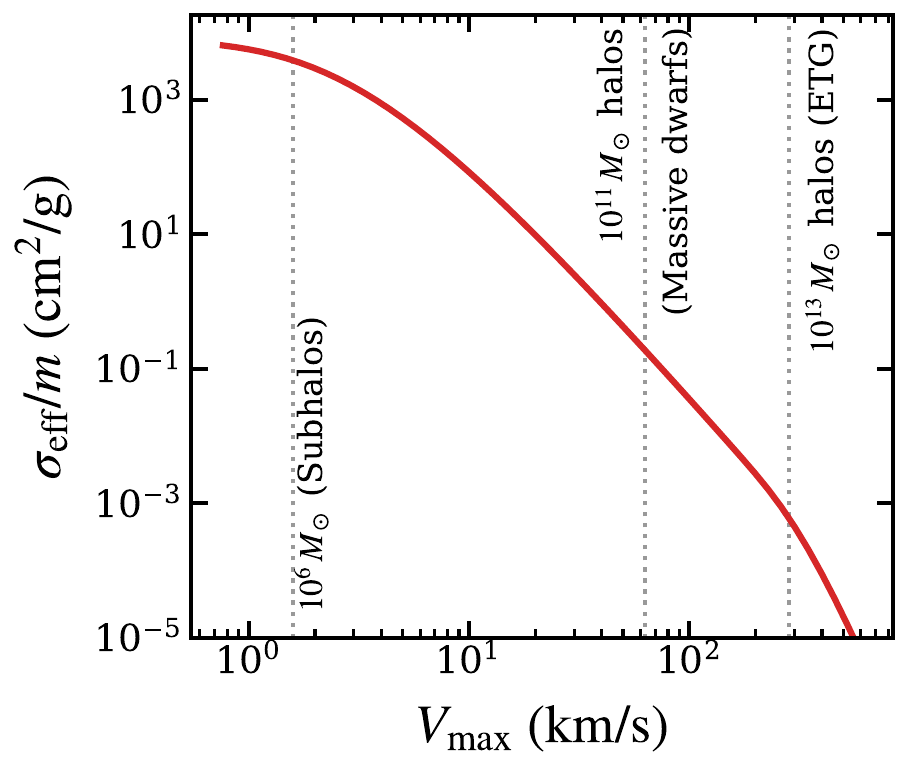}
    \includegraphics[width=0.518\linewidth]{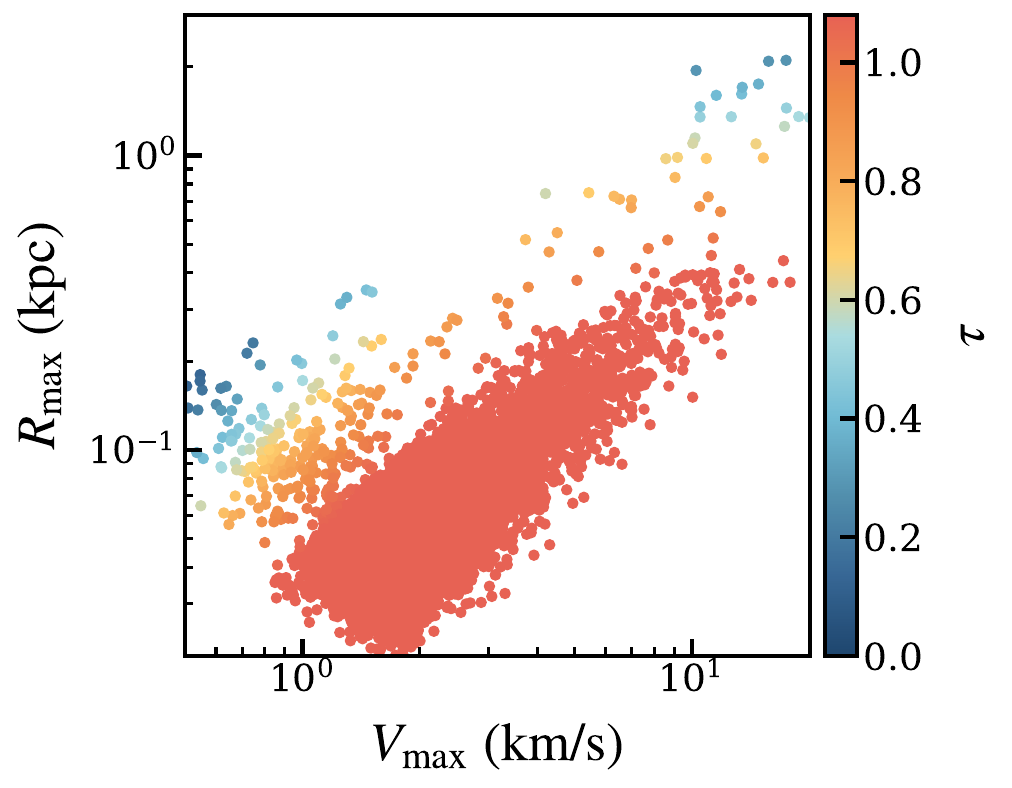}
    \caption{
        Velocity-dependent self-interaction and gravothermal evolution of SIDM halos.
        Left: the effective self-interaction cross section $\sigma_{\rm eff}/m$ as a function of $V_{\rm max}$, with vertical lines marking characteristic velocity scales for subhalos, massive dwarfs, and early-type galaxy (ETG) halos. The suppression at high velocities satisfies cluster constraints while allowing strong interactions on small scales.
        Right: SIDM halos in the $(V_{\rm max}, R_{\rm max})$ plane, colored by the gravothermal phase parameter $\tau$. Halos with $\tau \gtrsim 0.8$ are in the core-collapse regime, with most $M_h \lesssim 10^8,M_\odot$ halos occupying this region.
        }
        \label{fig:sigmaeff_vs_vmax}
\end{figure}

\EDFig{fig:sigmaeff_vs_vmax} (left) shows the effective constant SIDM cross section as a function of the maximum halo circular velocity, $V_{\max}$.
The effective cross section is obtained by velocity-averaging the underlying Rutherford-like, velocity-dependent differential scattering cross section.
The values of $V_{\max}$ are computed from rotation curves of halos with median concentration at the corresponding mass scales. In the SIDM scenario adopted here, the effective cross section decreases from $0.3$ to $0.035~\mathrm{cm^2\,g^{-1}}$ as the characteristic velocity increases from $55$ to $100~\mathrm{km\,s^{-1}}$.
Vertical dotted lines indicate the typical $V_{\max}$ values for subhalos, massive dwarf galaxies, and group- and cluster-scale halos, illustrating the effective SIDM strengths relevant for each regime.
At higher velocities corresponding to galaxy and cluster scales, the effective cross section is strongly suppressed, ensuring consistency with observational constraints from massive systems.
In contrast, at dwarf-galaxy and subhalo scales, the cross section remains large, enabling efficient heat transport and rapid gravothermal evolution.

%% The benchmark parameters used here are selected from a region constrained by independent astrophysical observations of dwarf-galaxy cores, Milky-Way analogues, and intermediate-mass halos. Within this allowed range, we adopt $\sigma_0/m = 8000\,\mathrm{cm^2\,g^{-1}}$ and $w = 6\,\mathrm{km\,s^{-1}}$, corresponding to an effective scattering cross-section of $\sigma_{\rm eff}/m \sim 0.3\,\mathrm{cm^2\,g^{-1}}$ for a $10^{11} M_\odot$ halo—consistent with~\RefCite{Zhang:2025bju} and representative of the center of the observationally permitted parameter band. For low-mass subhalos, the same scattering model predicts a steep rise in $\sigma_{\rm eff}(M)$, naturally placing them in the core-collapse regime. In the simulations, each subhalo is therefore assigned its mass-dependent effective cross-section, allowing the associated perturbations to emerge self-consistently from the SIDM microphysics rather than through imposed collapse.
To model the lensing signatures of subhalos, we adopt the analytic expressions for the deflection angle proposed in~\RefCite{Hou:2025gmv}, which describes SIDM halos. This model leverages the approximate universality of gravothermal evolution, which has been shown to apply to both isolated halos and subhalos. A physical interpretation of this universality is provided by introducing a dimensionless gravothermal phase, $\tau = t/t_c$, where $t$ is the halo's evolution time and $t_c$ is its core-collapse time~\cite{Outmezguine:2022bhq,Yang:2023jwn,Zhong:2023yzk}.

For dark matter–only halos, the core-collapse time can be written as~\cite{Koda:2011yb,Choquette:2018lvq,Yang:2023jwn}: 
\begin{equation}
    t_c = \frac{150}{C}\,
    \frac{1}{(\sigma/m)\,\rho_{s,0}\, r_{s,0}\, \sqrt{4\pi G \rho_{s,0}}},
\end{equation}
The normalization constant is fixed to $C=0.75$, which reproduces the corresponding trends seen in $N$-body simulations. Here $\rho_{s,0}$ and $r_{s,0}$ denote the NFW scale density and scale radius.

Using the NFW parameters $\rho_{s,0}$ and $r_{s,0}$ obtained for each subhalo from \texttt{pyHalo}, we compute the gravothermal phase $\tau$.
Once $\rho_{s,0}$, $r_{s,0}$, and $\tau$ are specified, the SIDM lensing model of~\RefCite{Hou:2025gmv} provides analytic expressions for the deflection angle.
We then combine the deflection angles of all subhalos with that of the host halo.
The magnification map is computed from the Jacobian determinant of the lens mapping.
This procedure yields a complete magnification map that incorporates SIDM effects.
It enables direct comparison with observational data and with predictions based on CDM.

%\EDFig{fig:sigmaeff_vs_vmax} shows the form of the effective SIDM cross section adopted in this work.
%It also illustrates the gravothermal evolution of SIDM halos across a wide range of mass scales under this cross section.
%As the characteristic velocity increases toward galaxy and cluster scales, the effective self-interaction cross section is strongly suppressed.
%This suppression ensures consistency with observational constraints from massive systems.
%In contrast, for dwarf galaxies and subhalos, the self-interaction cross section remains large.
%This enables efficient heat transport and drives rapid gravothermal evolution.

\EDFig{fig:sigmaeff_vs_vmax} (right) presents the distribution of subhalos and LOS halos in the CDM and SIDM scenarios for the lens systems shown in Fig.~\ref{fig:substructure_maps}, in the plane of halo mass and gravothermal evolutionary stage.
In particular, nearly all halos with $M_h \lesssim 10^8\,M_\odot$ satisfy $\tau \gtrsim 0.8$.This indicates that they have already entered the core-collapse phase.
Such velocity-dependent evolution ensures the collapse of a large population of low-mass halos.
It therefore provides a physical connection between velocity-dependent self-interactions and the emergence of rich small-scale features.

\subsubsection*{FDM}
In the FDM scenario, density fluctuations arising from quantum interference can be statistically described by the perturbation power spectrum, which sets the effective fluctuation amplitude projected onto the lens plane. The expected standard deviation of the convergence fluctuations is \cite{Laroche:2022pjm}
\begin{equation}
\sqrt{\langle \delta\kappa^2\rangle}
= 2.875 \times 0.025
\left(\frac{A_{\rm fluc}}{0.05}\right)
\left(\frac{f}{0.5}\right)
\left(\frac{m_\psi}{10^{-22}\,{\rm eV}}\right)^{-1/2}
\,\kappa_{\rm host}(r),\label{eq: fluctuations}
\end{equation}
where $A_{\rm fluc}$ denotes the dimensionless fluctuation amplitude calibrated from numerical simulations, $f$ is the dark matter mass fraction evaluated at the Einstein radius, with a typical value of $f \simeq 0.48 \pm 0.15$ \cite{Shajib:2020ywe}, and $\kappa_{\rm host}(\theta_E)$ represents the convergence of the host galaxy at the Einstein radius.

While \texttt{pyHalo} applies the fluctuation model at a single characteristic radius, FDM fluctuations are expected to vary with radius because they track the host-halo density. We therefore extend \texttt{pyHalo} by dividing the lens plane into concentric annuli. In each annulus, we evaluate $\kappa_{\rm host}(r)$ and set the fluctuation amplitude using \EDEq{eq: fluctuations} evaluated at that radius. This implementation reduces the fluctuation strength at larger radii, as expected from wave-interference simulations. 

We adopt a representative FDM benchmark with $m_\psi = 0.8\times10^{-22}\,\mathrm{eV}$ and a fluctuation amplitude calibrated from the simulations of~\RefCite{Schive:2014hza} ($A_{\mathrm{fluc}} = -1.3$) implemented following the prescription of~\RefCite{Chan:2020exg} in \texttt{pyHalo}. This benchmark provides a consistent forward-modeling template for assessing the impact of FDM-like fluctuations on the $R_{\rm cusp}$ plane. A full exploration of the $(m_\psi, A_{\rm fluc})$ parameter space would require dedicated field-level or wave-based simulations and is beyond the scope of this study. The adopted benchmark is sufficient for evaluating the scale and morphology of FDM fluctuations in comparison with CDM and SIDM scenarios.

Adopting the derivation of~\RefCite{Abe:2023hlf}, the convergence perturbation and the associated image-position perturbation are approximately linearly related: $\epsilon^2 \langle \delta\theta_x^2 \rangle = \tfrac{3}{2}\langle \delta\kappa^2\rangle$, where $\epsilon \sim 1/\theta_{\rm E}$. This relation implies that once the FDM-induced convergence fluctuation $\delta\kappa$ on the lens plane is specified, the corresponding image-plane displacement scale $\delta\theta$ follows directly. In the presence of FDM fluctuations, image positions are expected to deviate mildly from the smooth-lens prediction. We treat this as a physically motivated mapping that provides a reasonable positional tolerance for the MCMC sampling.

\subsection{Lensing effects of line of sight structures}
\label{Sec:Multi-plane lensing}
Structures along the LOS play a non-negligible role in strong lensing systems~\cite{Xu:2011ru, Li:2018cbz, Fleury:2021tke}. Thus, we employ a multi-plane lensing ray-tracing algorithm ~\cite{Schneider:1997bq, Li:2020fpq} to account for their influence. The multi-plane lensing equation in this work is given by~\cite{Gilman:2025fhy}
\begin{equation}
    \boldsymbol{\theta}_{n+1}
    =
    \boldsymbol{\theta}
    -
    \frac{1}{D_s}
    \sum_{i=1}^{n}
    D_{is}\,
    \boldsymbol{\alpha}_i\!\left(D_i \boldsymbol{\theta}_i\right),
\end{equation}
where $\boldsymbol{\theta}$ denotes the observed angular position, $\boldsymbol{\theta}_n$ represents the angular position of the light ray on the $n$-th lens plane, $\boldsymbol{\alpha}_i$ is the deflection field associated with the $i$-th lens plane, $D_s$ denotes the angular diameter distance to the source plane, and $D{ij}$ is the angular diameter distance between the $i$-th and $j$-th lens planes.
Hence, the total deflection angle can be written as
$
\boldsymbol{\alpha}_{\rm eff}
=
\frac{1}{D_s}
\sum_{i=1}^{n}
D_{is}\,
\boldsymbol{\alpha}_i\!\left(D_i \boldsymbol{\theta}_i\right).
$
The observed image therefore corresponds to viewing the source through a sequence of gravitational lenses along the line of sight. 
In this multi-plane framework, the effective convergence that summarizes the integrated lensing contribution can be calculated as
$\kappa_{\rm eff} \equiv \frac{1}{2}\nabla\!\cdot\!\boldsymbol{\alpha}_{\rm eff}$.

To isolate the lensing contribution from small-scale structures, we remove the mean convergence on each lens plane~\cite{Gilman:2025fhy}. 
For the $i$-th plane, the convergence is given by
$\kappa_i(\boldsymbol{\theta})=\frac{1}{2}\nabla\!\cdot\!\boldsymbol{\alpha}_i(\boldsymbol{\theta})$.
We then subtract its spatial mean to remove the uniform mass-sheet component and define
$\kappa_i'=\kappa_i-\bar{\kappa}_i$.
In practice, this is implemented by replacing $\boldsymbol{\alpha}_i$ with the renormalized deflection field $\boldsymbol{\alpha}_i'$. Subsequently, we emphasize the influence of substructures and LOS halos in Fig.~\ref{fig:substructure_maps} by defining

\begin{equation}
        \kappa_{\text{sub(eff)}} = \kappa_{\text{eff}} - \kappa_{\text{macro}},
        \label{eq:kappa_eff_sub}
\end{equation}
where $\kappa_{\text{macro}}$ denotes the contribution from the primary lens.

Through our calculations, we define the size of light cones as $b_{\text{sz}} \equiv 3\,\theta_{\text{E}}/\sqrt{q_{\text{SIE}}} $ and adopt an opening angle of $\theta_{\text{cone}} = 1.5\,b_{\text{sz}}$ to ensure that the simulated field of view consistently covers the region relevant for cusp configurations. Contributions from larger angular scales predominantly enter as smooth or slowly varying deflection fields, which are absorbed into the macromodel parameters, and do not induce localized distortions at the Einstein-radius scale \cite{Lin:2025dgn}.

Subhalos associated with the primary deflector are placed on a single lens plane at $z_l$, following a projected NFW distribution. LOS halos are populated along the light cone and assigned to discrete planes using the default \texttt{pyHalo} spacing of $\Delta z = 0.02$. Halo masses are sampled from $M = 10^{-7.5} M_{\text{host}}$ to $10^{-1} M_{\text{host}}$. All lensing calculations are performed on a \( 1500 \times 1500 \) grid.

%\subsection*{MCMC Matching Procedures for Each $\phi$ Stage}
\subsection{MCMC Sampling of Cusp Configurations}
\label{sec:mcmc_sampling}

The caustic cross section for cusp configurations is very small, thus it is time-consuming to conduct naive random sampling of the source plane. To achieve adequate coverage of cusp-like quadruply imaged systems, 
We employ a two-stage MCMC sampling scheme in the source plane, 
implemented using the \texttt{emcee} package~\cite{Foreman-Mackey:2012any}\footnote{\url{https://github.com/dfm/emcee}}.

Given a fixed lens model, the MCMC samples $N$ positions $\{(x_s^{(i)}, y_s^{(i)})\}_{i=1}^{N}$ on source plane and then we map it to four-image configurations in lens plane. 
Our goal is to retain those configurations whose image geometry (characterized by $\phi$, $\phi_1$, and $\phi_2$) is consistent with systems exhibiting a cusp configuration, as shown in~\EDFig{fig:Major_Minor}.

For a given source position, we compute the opening angles $\phi$ by mapping the source-plane coordinates $(x_s, y_s)$ to the corresponding image-plane positions $\{(x_l^{(j)}, y_l^{(j)})\}_{j=1}^{4}$. 
The method proceeds via the following steps:
\begin{enumerate}
\item[i.] The lens equation is sampled on a regular grid in the image plane, with each grid point mapped back to the source plane. These mapped points are connected to form a triangular mesh in the source plane.
\item[ii.] For a given source position, the enclosing triangle in this mesh is identified. 

\item[iii.] Because the mapping from the image plane to the source plane is generally many-to-one, a single source position $(x_s, y_s)$ may correspond to multiple image-plane locations. We therefore establish the mapping from $(x_s, y_s)$ to each relevant image-plane triangle using barycentric interpolation (see Fig.~3 of \cite{Li:2020fpq}).
\item[iv.] The magnifications at these locations are obtained by evaluating the magnification map of this lens model.

\end{enumerate}
From the full solution set, the three images with the highest magnifications $\mu^{(j)}$ are identified as the cusp triplet.
The angles $\phi$, $\phi_1$, and $\phi_2$ for this triplet are then computed with respect to the lens center, as schematically illustrated in \EDFig{fig:Major_Minor}.

%This is achieved by first sampling the lens equation on a regular grid in the image plane and mapping each grid \mkred{position} to the source plane. 
%The resulting mapping is then divided into a triangular mesh in the source plane, where each triangle is paired with its corresponding triangle in the image plane.

%This yields the corresponding image positions, while the magnifications are obtained by interpolating the magnification map at these locations. Repeating this procedure over all intersected triangles provides the full set of image solutions.

%Finally, the magnification $\mu^{(j)}$ of each image is obtained by interpolating the magnification map at the corresponding image-plane positions $\{(x_l^{(j)}, y_l^{(j)})\}_{j=1}^{4}$. We define the cusp triplet as the three images with the highest magnifications, from which the angles $\phi$, $\phi_1$, and $\phi_2$ are computed with respect to the lens center.

For each individual source realization, we construct a likelihood by comparing
the resulting image geometry with a target cusp configuration defined within a given opening-angle bin.
Specifically, the misfit is quantified by
%The misfit is quantified by
\begin{equation}
\chi_k^{2}(x_s,y_s)
=
\left(\frac{\phi-\phi_{\rm targ}^{(k)}}{\sigma_{\phi}}\right)^{2}
+
\left(\frac{\phi_{1}-\phi_{1,{\rm targ}}^{(k)}}{\sigma_{\phi_{1}}}\right)^{2},
\label{Eq:kasquare}
\end{equation}
where $\phi_{\rm targ}^{(k)}$ denotes the central value of the $k$-th opening-angle bin, while we require $\phi_{1,{\rm targ}}^{(k)} = \phi_{\rm targ}^{(k)}/2$ for the cusp geometry.
The parameters $\sigma_{\phi}$ and $\sigma_{\phi_1}$ represent the tolerated angular widths that define the acceptance region in the $(\phi,\phi_1)$ space. For conservative reasons, we adopt the FDM-induced fluctuation scale to define $\delta\phi_{\rm c}$, as FDM produces larger perturbations than CDM or SIDM.
In Fig.~\ref{fig:substructure_maps}, $\delta\phi_{\rm c}\simeq 0.01^\circ$, with comparable values for other lenses, well below the $\phi$-bin width.

By scanning over a sequence of opening-angle bins indexed by $k$ and minimizing $\chi_k^{2}$ within each bin, we efficiently sample source positions that realize the desired cusp configurations.

%The posterior distribution for the source position is then
%\begin{equation}
%p(x_s,y_s\,|\,{\rm data}) \propto \exp\!\left[-\frac{1}{2}\chi^{2}(x_s,y_s)\right].
%\end{equation}
%In practice, \texttt{emcee} evolves an ensemble of walkers in the $(x_s, y_s)$ plane to sample the posterior over source positions that reproduce the targeted cusp geometry. The walkers are initialized near the lens center and are then iteratively updated according to the posterior weight set by the likelihood. Specifically,
Our two-stage MCMC procedure is described as follows. Using \texttt{emcee}, we initialize the source positions as
\begin{equation}
(x_{s,0}^{(i)},y_{s,0}^{(i)}) = (0,0) + \sigma_{\rm init}\,\boldsymbol{\eta}^{(i)},
\qquad
\boldsymbol{\eta}^{(i)} \sim \mathcal{N}(\mathbf{0},\mathbf{I}),
\qquad
i=1,\ldots,n_{\rm walkers},
\end{equation}
where $\sigma_{\rm init}$ controls the initial scatter and $n_{\rm walkers}$ is the number of walkers.
The origin $(0,0)$ is set at the lens center corresponding to the line of sight, while all selection is imposed through the $\chi_k^{2}(\phi,\phi_1)$ constraint.

The first coarse stage adopts a step size of $r_{\rm eff}/8$. The characteristic sampling scale $r_{\rm eff}$ is defined as $\sqrt{A_c/\pi}$, where $A_c$ denotes the area enclosed by the tangential caustic of the smooth model.
Five independent chains of 600 steps are executed to efficiently identify regions with consistent image geometries. 
The resulting $\chi^2$ minimum seeds a refined second stage, 
which adopts a tighter tolerance ($\delta\phi_{\rm f} = \delta\theta/2$) and 
smaller step size $r_{\rm eff}/16$ among ten 1000-step chains to ensure convergence. 
We confirm that all chains converge within 300 steps.

To ensure uniform source-plane coverage, we implement a grid-based resampling procedure. 
Samples exceeding a $5\sigma$ threshold are discarded, and the remainder are projected onto a $100 \times 100$ grid. 
We require only the minimum-$\chi^2$ sample within each grid and assign it a normalized weight $w \propto \exp(-\chi^2/2)$.

%%%%%%%%%%%%%%%%%%%%%%%%%%%%%%%%%%%%%%%%%%%%%%%%%%%%%%%%
\subsection{Theoretical Prediction Maps}
%%%%%%%%%%%%%%%%%%%%%%%%%%%%%%%%%%%%%%%%%%%%%%%%%%%%%%%%

%To construct theoretical prediction maps in the $R_{\rm cusp}$ plane, we generate $10^8$ points for the following scenarios: a smooth model and realizations including CDM, SIDM, and FDM perturbations. Our sample consists of 572 mock lenses, comprising 324 SIE and 248 multipole models. 
Theoretical predictions in the $R_{\rm cusp}$ plane are generated using a sample of 572 mock lenses (324 SIE models and 248 multipole models). We calculate $\sim10^8$ data points (i.e. $10^4$ grid points, $12$ $\phi$ bins, $572$ mock lenses and $4$ scenarios) for two cases: 
a smooth analytical model and simulations that include the perturbations induced by CDM, SIDM, and FDM.
For each lens, we divide the opening angle range $30^\circ<\phi<150^\circ$ into twelve bins with bin size $10^\circ$.
%with $10^\circ$ centers and $\pm 5^\circ$ half-widths.

Within each bin, we identify source positions $(x_s, y_s)$ whose mapped images satisfy the cusp-geometry requirements. 
For each accepted position, we solve the multi-plane lens equation to determine image magnifications. 
The $R_{\rm cusp}$ value is then computed from the three most highly magnified images (the cusp triplet). 
This procedure yields $\sim 10^4$ samples per $\phi$ bin, providing a robust statistical basis for comparing dark matter scenarios. Finally, we obtain the weighted set $\{(R_{\rm cusp}^{(i,k)}, w_k^{(i,k)})\}_{i=1}^{N_k}$ for the $k$-th $\phi$ bin.

%%%%%%%%%%%%%%%%%%%%%%%%%%%%%%%%%%%%%%%%%%%%%%%%%%%%%%%%
\subsection{Bayesian Decision Criterion}
%%%%%%%%%%%%%%%%%%%%%%%%%%%%%%%%%%%%%%%%%%%%%%%%%%%%%%%%
\label{Sec:Bey}
The resulting $R_{\rm cusp}$ distribution in each $\phi$ bin is calculated using a kernel density estimate that is normalized independently within that bin. 
We compare cusp configurations at the same $\phi$. 
%This strategy removes ambiguities introduced by changes in caustic geometry when additional structure is included in the lens model. 
As a result, differences in the $R_{\rm cusp}$ distributions provide a direct measurement of small-scale perturbations.
To quantitatively assess the consistency between model predictions and observations, we adopt Bayes factor. 
For each candidate model $M$, the evidence for a fixed $\phi$ is
\begin{equation}
Z_M \propto \int p(R_{\rm cusp} \mid M)\,\mathcal{L}_{\rm obs}(R_{\rm cusp})\, dR_{\rm cusp},
\end{equation}
where $R_{\rm obs}$ is the observed value of $R_{\rm cusp}$ with uncertainty $\sigma_{\rm obs}$. 
The model-predicted prior is $p(R_{\rm cusp} \mid M)$, and $\mathcal{L}_{\rm obs}$ denotes the Gaussian likelihood, given by
\begin{equation}
\mathcal{L}_{\rm obs}(R_{\rm cusp}) =
\frac{1}{\sqrt{2\pi}\sigma_{\rm obs}}
\exp\!\left[
-\frac{(R_{\rm cusp}-R_{\rm obs})^{2}}{2\sigma_{\rm obs}^{2}}
\right].
\end{equation}

For multiple independent measurements, we combine their evidences. For a given model $M$, the joint evidence is therefore calculated as $Z_M^{\rm tot} = \prod_{s} Z_{M,s}$. Here $Z_{M,s}$ denotes the evidence evaluated for the $s$-th observational systems. Accordingly, we compare models by computing the combined Bayes factor between two models $M_i$ and $M_j$ can be written as
\begin{equation}
\mathrm{BF}_{ij}
= \frac{Z_{M_i}^{\rm tot}}{Z_{M_j}^{\rm tot}}
= \prod_{s} \frac{Z_{M_i,s}}{Z_{M_j,s}}
= \prod_{s} \mathrm{BF}_{ij,s}.
\end{equation}

\bmhead{Acknowledgements}

We thank Daniel Gilman, Amruth Alfred, Jeremy Lim, Dandan Xu, Massimo Meneghetti, Wenwen Zheng, Han Qu, and Lei Wang for helpful discussions.
G.L. acknowledges support from the National Natural Science Foundation of China (NSFC) (No. 12533008). 
D.Y., Y.F. and Y.-L.S.T. were supported in part by the National Key Research and Development Program of China (No. 2022YFF0503304), Natural Science Foundation of China under No. 12588101, Chinese Academy of Sciences (CAS), and the Project for Young Scientists in Basic Research of CAS (No. YSBR-092).
Y.S. acknowledges the support from the China Manned Space Program with grant no. CMS-CSST-2025-A20 and the National Natural Science Foundation of China (Grant No. 12333001).
N.L. acknowledges support from the CAS Project for Young Scientists in Basic Research (No. YSBR-062).
Z.H. acknowledges support from the National Natural Science Foundation of China (Grant No.~12403104).

\bmhead{Author contributions}

D.Y., Y.F. and Y.-L.S.T. proposed the general scientific direction of the study.
S.H. performed the full analysis, wrote the manuscript, and produced all figures. 
S.H. and S.X. jointly compiled photometric measurements and observational parameters for the cusp lens systems from the literature. For systems or quantities not available in published works, S.X. derived the required measurements using lens-modelling software. 
Y.-L.S.T. contributed to revising the main text, proposed the MCMC-based sampling strategy, and provided critical input on the MCMC and Bayesian analysis sections; these components were implemented by S.H. 
D.Y. provided guidance on the interpretation and modelling of dark matter substructure in the SIDM framework; the corresponding analysis was implemented by S.H. 
Y.S. provided guidance on the observational aspects of the study and commented on the manuscript. 
N.L. provided guidance on the implementation of lensing simulations, which were carried out by S.H. 
Y.F. and G.L. contributed to the overall scientific framework of the paper. 
Z.H. provided the initial results for all currently observed lensed quasar systems, which were subsequently selected and analyzed by S.H.
J.D. provided mock data based on SIE plus external shear models, which were extended by S.H. to include multipoles. 
All authors commented on the manuscript and approved the final version.

\bmhead{Competing interests} 
The authors declare no competing interests.

\bmhead{Data availability}
The data underlying the figures and main findings of this study are publicly available at
\url{https://github.com/HouSiyuan2001/CuspDM}.

\bmhead{Code availability}
The code used to generate the data and figures in this study is publicly available at
\url{https://github.com/HouSiyuan2001/CuspDM}.

\section*{Supplementary Information}
\bmhead{Impact of axis separation on model discrimination}

\EDTable{tab:bf_combined} presents the Bayesian model comparison for the combined cusp sample, without separating major- and minor-axis configurations.
In contrast to the axis-separated analysis, the Bayes factors between different dark-matter scenarios are substantially reduced, typically being of order unity. 
This degradation in discriminating power reflects the fact that, at the same opening angle $\phi$, minor-axis configurations produce images closer to the critical curve.
In addition, the corresponding sources lie closer to the cusp, yielding systematically lower $R_{\rm cusp}$ values and enhanced sensitivity to dark-matter substructure.
When these configurations are combined, this geometric distinction is erased, since the larger $R_{\rm cusp}$ values from major-axis configurations dominate at fixed $\phi$. 
As a result, the excess signal from the minor-axis cases is diluted, reducing the overall sensitivity to dark-matter substructure.

\bmhead{Limited discriminating power of major-axis cusps at large opening angles}

\EDTable{tab:bf_major_phi70} shows the Bayesian model comparison for major-axis cusp configurations with opening angles $\phi>70^\circ$. In this regime, the Bayes factors across different dark-matter scenarios are close to unity. 
This reflects the low sensitivity of large-$\phi$ major-axis cusps, whose images lie far from the critical curve and are therefore weakly affected by dark-matter substructure. In this regime, larger $R_{\rm cusp}$ responses predominantly arise from the intrinsic macromodel geometry rather than from perturbations. As a result, future observations targeting minor-axis cusps or narrow major-axis cusps with $\phi \lesssim 70^\circ$ are expected to provide significantly greater discriminatory power between different dark-matter models.

\bmhead{Robustness to microlensing-free subsamples}

\EDTable{tab:bf_matrix_full} presents the Bayesian model comparison after excluding systems with flux ratios measured in the \textit{HST} F160W band, thereby restricting the analysis to subsamples widely regarded as microlensing-free. The resulting Bayes factors are consistent with those obtained from the full sample: smooth lens models remain decisively disfavoured, and FDM is strongly preferred over CDM and SIDM, both with and without multipole freedom. The persistence of these results demonstrates that our main conclusions are not driven by systems potentially affected by microlensing.
\begin{table*}[htbp]
\centering
\caption{
Key parameters of the mock lens system in Fig.~\ref{fig:substructure_maps}.
The table lists the parameters that define the lens macromodel and its higher-order multipole components. The macromodel is described by an \texttt{SIE} mass profile, with external shear and higher-order angular structures included via the \texttt{SHEAR} and \texttt{MULTIPOLE\_ELL} models in \texttt{lenstronomy}. Here $a$ denotes the semi-major axis of the ellipse~\cite{Oh:2024dlj}, which reduces to $a=\theta_E/\sqrt{q_{\rm SIE}}$.
}
\label{tab:mock_all_sim_185_key}
\begin{tabular}{cccccc}
\hline
$z_l$ & $z_s$ & $\theta_E\,(\mathrm{arcsec})$ & $q_{\rm SIE}$ & $\gamma_{\rm ext}$ &  \\
\hline
1.1471 & 3.7039 & 0.5261 & 0.4116 & 0.1261 &  \\
\hline\hline
$\phi_{\rm ext}\,(\mathrm{rad})$ & $a_3/a$ & $\Delta\phi_{m3}\,(\mathrm{rad})$ & $a_4/a$ & $\Delta\phi_{m4}\,(\mathrm{rad})$ &  \\
\hline
3.6360 & 0.00623 & $-0.4230$ & 0.06350 & $-0.0933$ &  \\
\hline
\end{tabular}
\end{table*}

\begin{figure}
    \centering
    \includegraphics[width=0.7\linewidth]{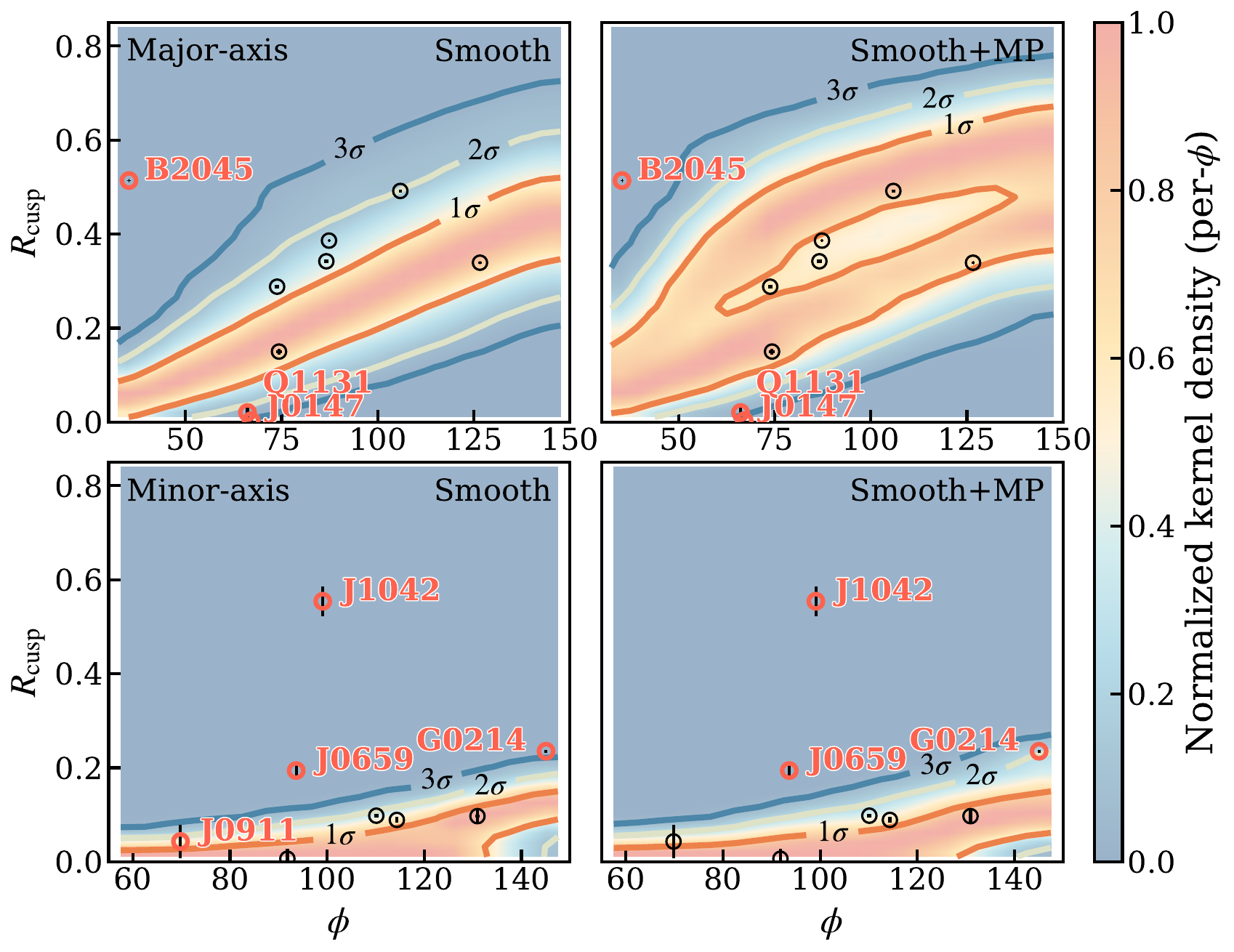}
    \caption{Same as Fig.~\ref{fig:NomonizedKDE}, but for smooth lens.}
     \label{fig:NomonizedKDE_MP_Smooth}
\end{figure}
 \begin{table}
    \centering
    \footnotesize
    \caption{
    The same as Table~\ref{tab:Bey}, but for the Smooth and Smooth+MP models.
    }
    \label{tab:Bey_smooth}
    \begin{tabular}{lcccccc}
    \toprule
    \multicolumn{7}{c}{
    $\mathrm{BF(row)/BF(column)}$
    } \\
    \midrule
            & CDM & SIDM & CDM+MP & SIDM+MP & FDM+MP & FDM \\
    \midrule
    Smooth
            & $10^{-68}$
            & $10^{-71}$
            & $10^{-71}$
            & $10^{-72}$
            & $10^{-73}$
            & $10^{-75}$ \\
    Smooth+MP
            & $10^{-35}$
            & $10^{-37}$
            & $10^{-38}$
            & $10^{-38}$
            & $10^{-39}$
            & $10^{-41}$ \\
    \bottomrule
    \end{tabular}
\end{table}
\begin{table}[t]
\centering
\footnotesize
\caption{Same as Table~\ref{tab:Bey} and \EDTable{tab:Bey_smooth}, but for the combined sample without separating major- and minor-axis cusp configurations.}
\label{tab:bf_combined}
\begin{tabular}{lccccccc}
\toprule
& \multicolumn{7}{c}{$\mathrm{BF}(\mathrm{row})/\mathrm{BF}(\mathrm{column})$} \\
\midrule
& Smooth & Smooth+MP & CDM & SIDM & CDM+MP & SIDM+MP & FDM+MP \\
\midrule
Smooth+MP & 4.5 \\
CDM       & $10^{4}$ & 815 \\
SIDM      & $10^{4}$ & $10^{3}$ & 1.8 \\
CDM+MP    & $10^{4}$ & $10^{3}$ & 1.9 & 1.1 \\
SIDM+MP   & $10^{4}$ & $10^{3}$ & 2.2 & 1.2 & 1.1 \\
FDM+MP    & $10^{3}$ & 430 & 0.53 & 0.29 & 0.27 & 0.24 \\
FDM       & $10^{5}$ & $10^{4}$ & 34 & 19 & 18 & 16 & 64 \\
\bottomrule
\end{tabular}
\end{table}

\begin{table}[t]
\centering
\footnotesize
\caption{Same as Table~\ref{tab:Bey} and \EDTable{tab:Bey_smooth}, but restricting to major-axis cusp configurations with $\phi>70^\circ$.}
\label{tab:bf_major_phi70}
\begin{tabular}{lccccccc}
\toprule
& \multicolumn{7}{c}{$\mathrm{BF}(\mathrm{row})/\mathrm{BF}(\mathrm{column})$} \\
\midrule
& Smooth & Smooth+MP & CDM & SIDM & CDM+MP & SIDM+MP & FDM+MP \\
\midrule
Smooth+MP & 2.0 \\
CDM       & 1.1 & 0.55 \\
SIDM      & 1.5 & 0.72 & 1.3 \\
CDM+MP    & 2.3 & 1.1  & 2.0 & 1.6 \\
SIDM+MP   & 2.3 & 1.1  & 2.0 & 1.5 & 0.99 \\
FDM+MP    & 1.6 & 0.78 & 1.4 & 1.1 & 0.69 & 0.70 \\
FDM       & 7.8 & 3.9  & 7.0 & 5.3 & 3.4  & 3.4 & 4.9 \\
\bottomrule
\end{tabular}
\end{table}

\begin{table}[t]
\centering
\footnotesize
\caption{Same as Table~\ref{tab:Bey} and \EDTable{tab:Bey_smooth}, but excluding systems whose flux ratios are measured in the \textit{HST} F160W band.}
\label{tab:bf_matrix_full}
\begin{tabular}{lccccccc}
\toprule
& \multicolumn{7}{c}{$\mathrm{BF}(\mathrm{row})/\mathrm{BF}(\mathrm{column})$} \\
\midrule
& Smooth & Smooth+MP & CDM & SIDM & CDM+MP & SIDM+MP & FDM+MP \\
\midrule
Smooth+MP  & $10^{26}$ \\
CDM        & $10^{61}$ & $10^{35}$ \\
SIDM       & $10^{64}$ & $10^{37}$ & 356 \\
CDM+MP     & $10^{64}$ & $10^{37}$ & 319 & 0.9 \\
SIDM+MP    & $10^{64}$ & $10^{38}$ & $10^{3}$ & 4.3 & 4.8 \\
FDM+MP     & $10^{65}$ & $10^{39}$ & $10^{4}$ & 26 & 29 & 6.0 \\
FDM        & $10^{66}$ & $10^{40}$ & $10^{5}$ & 395 & 441 & 92 & 15 \\
\bottomrule
\end{tabular}
\end{table}
\begin{table}[t]
\centering
\footnotesize
\caption{Normalized per-$\phi$ kernel densities shown in Fig.~\ref{fig:NomonizedKDE}, 
Fig.~\ref{fig:NomonizedKDE_MP}, and \EDFig{fig:NomonizedKDE_MP_Smooth} for individual cusp lenses.
Shading corresponds to 1--2$\sigma$ (blue), 2--3$\sigma$ (yellow), and $>3\sigma$ (red),
while unshaded cells lie within $1\sigma$.}
\label{tab:individual_norms}
\setlength{\tabcolsep}{6pt}
\renewcommand{\arraystretch}{1.15}

\begin{tabular}{lcccccccc}
\toprule
System 
& Smooth 
& Smooth+MP 
& CDM 
& SIDM 
& CDM+MP 
& SIDM+MP 
& FDM+MP 
& FDM \\
\midrule

\multicolumn{9}{c}{Major-axis systems} \\
\midrule

B2045
& \cellcolor{sigOut!50}$10^{-7}$ 
& \cellcolor{sigOut!50}$10^{-5}$ 
& \cellcolor{sigOut!50}$0.011$ 
& \cellcolor{sigTwoThree!50}$0.015$ 
& \cellcolor{sigOut!50}$0.0062$ 
& \cellcolor{sigOut!50}$0.012$ 
& \cellcolor{sigOneTwo!50}$0.21$ 
& \cellcolor{sigOneTwo!50}$0.37$ \\

Q1131
& \cellcolor{sigTwoThree!50}$0.062$ 
& \cellcolor{sigTwoThree!50}$0.059$ 
& \cellcolor{sigOneTwo!50}$0.24$ 
& \cellcolor{sigOneTwo!50}$0.28$ 
& \cellcolor{sigTwoThree!50}$0.19$ 
& \cellcolor{sigOneTwo!50}$0.28$ 
& \cellcolor{sigOneTwo!50}$0.32$ 
& \cellcolor{sigOneTwo!50}$0.55$ \\

J0147
& \cellcolor{sigOut!50}$0.016$ 
& \cellcolor{sigOut!50}$0.012$ 
& \cellcolor{sigOneTwo!50}$0.20$ 
& \cellcolor{sigOneTwo!50}$0.25$ 
& \cellcolor{sigTwoThree!50}$0.13$ 
& \cellcolor{sigOneTwo!50}$0.23$ 
& \cellcolor{sigOneTwo!50}$0.31$ 
& \cellcolor{sigOneTwo!50}$0.54$ \\

J0630
& \cellcolor{sigOneTwo!50}$0.31$ 
& \cellcolor{sigOneTwo!50}$0.58$ 
& \cellcolor{sigOneTwo!50}$0.43$ 
& \cellcolor{sigOneTwo!50}$0.46$ 
& \cellcolor{sigOneTwo!50}$0.56$ 
& \cellcolor{sigOneTwo!50}$0.60$ 
& $0.75$ 
& $0.94$ \\

B1422
& $0.93$ 
& $0.94$ 
& $0.91$ 
& $0.94$ 
& $0.83$ 
& $0.76$ 
& \cellcolor{sigOneTwo!50}$0.45$ 
& $0.75$ \\

J0833
& \cellcolor{sigOneTwo!50}$0.35$ 
& \cellcolor{sigOneTwo!50}$0.52$ 
& \cellcolor{sigOneTwo!50}$0.42$ 
& \cellcolor{sigOneTwo!50}$0.43$ 
& \cellcolor{sigOneTwo!50}$0.53$ 
& \cellcolor{sigOneTwo!50}$0.57$ 
& $0.77$ 
& $0.9$ \\

J0429
& \cellcolor{sigOneTwo!50}$0.22$ 
& \cellcolor{sigOneTwo!50}$0.57$ 
& \cellcolor{sigOneTwo!50}$0.27$ 
& \cellcolor{sigOneTwo!50}$0.28$ 
& \cellcolor{sigOneTwo!50}$0.59$ 
& \cellcolor{sigOneTwo!50}$0.62$ 
& $0.84$ 
& $0.66$ \\

B1608
& \cellcolor{sigOneTwo!50}$0.13$ 
& $0.81$ 
& \cellcolor{sigOneTwo!50}$0.15$ 
& \cellcolor{sigOneTwo!50}$0.15$ 
& $0.78$ 
& $0.81$ 
& $0.89$ 
& \cellcolor{sigOneTwo!50}$0.34$ \\

J1433
& $0.79$ 
& $0.64$ 
& $0.77$ 
& $0.80$ 
& $0.70$ 
& $0.76$ 
& \cellcolor{sigOneTwo!50}$0.51$ 
& $0.76$ \\

\midrule
\multicolumn{9}{c}{Minor-axis systems} \\
\midrule

J0911
& \cellcolor{sigTwoThree!50}$0.19$ 
& \cellcolor{sigOneTwo!50}$0.29$ 
& \cellcolor{sigOneTwo!50}$0.52$ 
& $0.62$ 
& $0.58$ 
& $0.67$ 
& $0.95$ 
& $0.98$ \\

J1251
& $1.00$ & $1.00$ & $1.00$ & $1.00$ & $1.00$ & $1.00$ & $0.98$ & $1.00$ \\

J0659
& \cellcolor{sigOut!50}$0.00$ 
& \cellcolor{sigOut!50}$10^{-6}$ 
& \cellcolor{sigTwoThree!50}$0.028$ 
& \cellcolor{sigTwoThree!50}$0.041$ 
& \cellcolor{sigOneTwo!50}$0.089$ 
& \cellcolor{sigOneTwo!50}$0.09$ 
& $0.89$ 
& $0.57$ \\

J1042
& \cellcolor{sigOut!50}$10^{-10}$ 
& \cellcolor{sigOut!50}$10^{-9}$ 
& \cellcolor{sigOut!50}$10^{-6}$ 
& \cellcolor{sigOut!50}$10^{-4}$ 
& \cellcolor{sigOut!50}$0.0038$ 
& \cellcolor{sigOut!50}$0.0057$ 
& \cellcolor{sigTwoThree!50}$0.14$ 
& \cellcolor{sigTwoThree!50}$0.061$ \\

J0029
& \cellcolor{sigOneTwo!50}$0.42$ 
& \cellcolor{sigOneTwo!50}$0.41$ 
& \cellcolor{sigOneTwo!50}$0.55$ 
& $0.59$ 
& $0.59$ 
& $0.63$ 
& $0.93$ 
& $0.80$ \\

J2205
& \cellcolor{sigOneTwo!50}$0.42$ 
& \cellcolor{sigOneTwo!50}$0.41$ 
& \cellcolor{sigOneTwo!50}$0.55$ 
& $0.59$ 
& $0.59$ 
& $0.63$ 
& $0.93$ 
& $0.80$ \\

J2017
& $1.00$ 
& $0.93$ 
& $0.88$ 
& $0.88$ 
& $1.00$ 
& $0.97$ 
& $0.93$ 
& $0.91$ \\

G0214
& \cellcolor{sigOut!50}$0.0091$ 
& \cellcolor{sigOneTwo!50}$0.20$ 
& \cellcolor{sigOneTwo!50}$0.086$ 
& \cellcolor{sigTwoThree!50}$0.073$ 
& \cellcolor{sigOneTwo!50}$0.30$ 
& \cellcolor{sigOneTwo!50}$0.31$ 
& \cellcolor{sigOneTwo!50}$0.62$ 
& $0.66$ \\

\bottomrule
\end{tabular}
\end{table}

\clearpage
\bibliography{sn-bibliography}% common bib file

\end{document}